%% file: EMC-RC.tex
\documentclass[aps,prl,superscriptaddress,showpacs,twocolumn,floatfix,amsmath,amssymb]{revtex4-2}

\usepackage{hyperref} 
\usepackage[mathlines]{lineno} 
\usepackage{graphicx} 
\usepackage{xcolor}
\usepackage{booktabs}
\usepackage{multirow}
\usepackage{array}

\def\gtorder{\mathrel{\raise.3ex\hbox{$>$}\mkern-14mu
 \lower0.6ex\hbox{$\sim$}}}
\def\ltorder{\mathrel{\raise.3ex\hbox{$<$}\mkern-14mu
 \lower0.6ex\hbox{$\sim$}}}

\begin{document}

\title{On the significance of radiative corrections on measurements of the EMC effect}
\input{authors.tex}

\date{\today}

\begin{abstract}
Analyzing global data on the EMC effect, which denotes differences in parton distribution functions in nuclei compared to unbound nucleons, reveals tensions. Precise measurements at Jefferson Lab, studying both $x$ and A dependence, show systematic discrepancies among experiments, making the extraction of the A dependence of the EMC effect sensitive to the selection of datasets. By comparing various methods and assumptions used to calculate radiative corrections, we have identified differences that, while not large, significantly impact the EMC ratios and show that using a consistent radiative correction procedure resolves this discrepancy, leading to a more coherent global picture, and allowing for a more robust extraction of the EMC effect for infinite nuclear matter.

\end{abstract}

\maketitle

\section{Introduction}

Precision measurement of deep-inelastic scattering (DIS) from nucleons and nuclei allows for an extraction of parton distribution functions (pdfs). Early measurements of the nuclear pdfs demonstrated that the quark distributions in iron were not simply the sum of the distribution arising from its constituent protons and neutrons~\cite{EuropeanMuon:1983wih}. Since the initial observation of this ``EMC effect'', additional measurements have been made for a wide range of nuclei~\cite{Gomez:1993ri, Dasu:1993vk, Seely:2009gt, CLAS:2019vsb, Arrington:2021vuu, JeffersonLabHallATritium:2021usd, HallC:2023xkj} and many different explanations have been proposed to explain this observation~\cite{Malace:2014uea, Arrington:2015wja, Hen:2016kwk, Cloet:2019mql, Arrington:2021vuu, Arrington:2022sov}. One of these measurements~\cite{Seely:2009gt} demonstrated an unexpected, non-trivial dependence on nuclear structure in light nuclei~\cite{Seely:2009gt} which corresponded to a similar dependence in the number of short-range correlations (SRCs)~\cite{Fomin:2011ng, Arrington:2012ax}, renewing interest in precision measurements of the A dependence of the EMC effect.

\begin{figure}[htb]
\begin{center}
\includegraphics[width=0.95\columnwidth,height=6.1cm,trim={3mm 0mm 16mm 18mm}, clip]{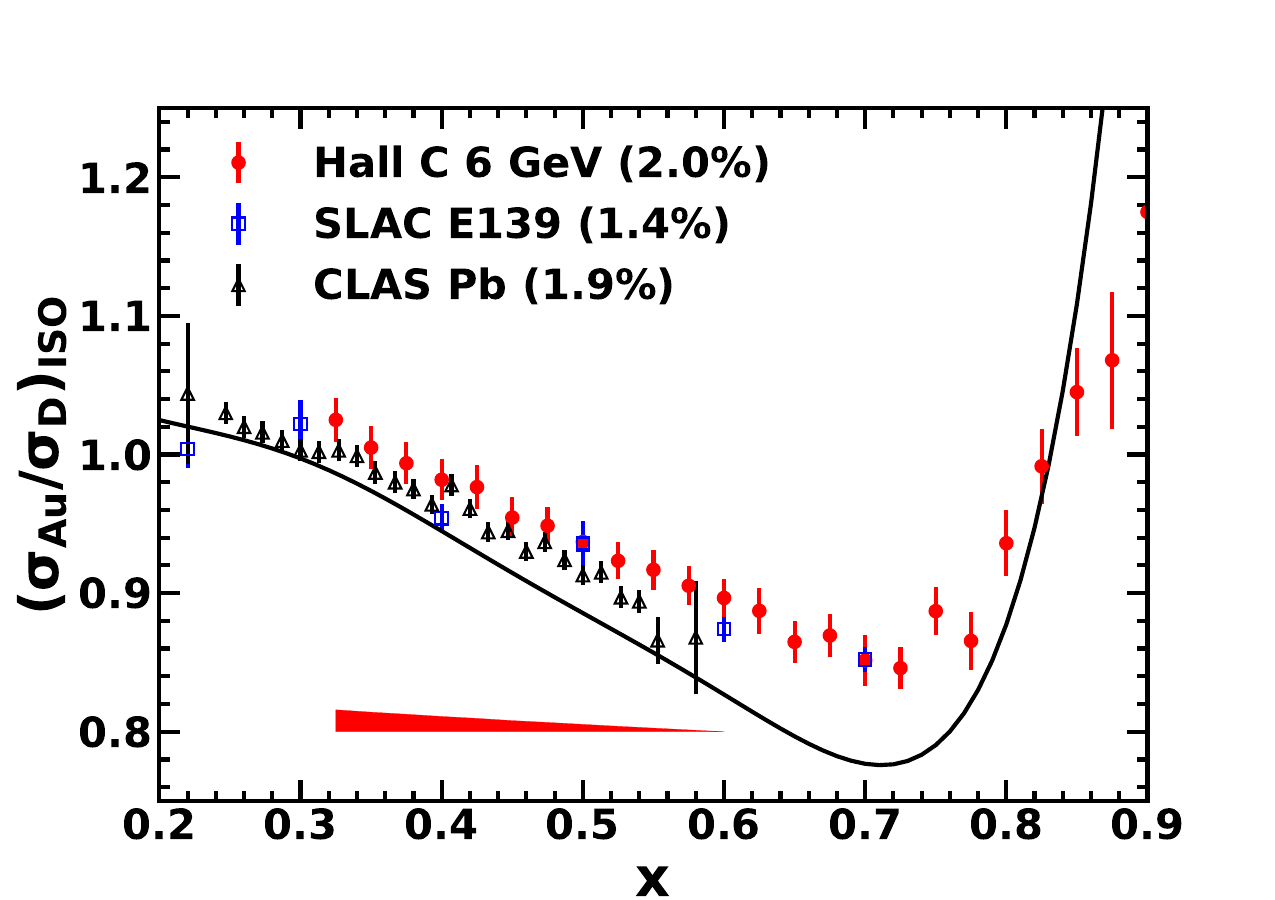}
\caption{EMC ratios for gold and lead~\cite{Gomez:1993ri, CLAS:2019vsb, Arrington:2021vuu} after applying uniform isoscalar corrections from~\cite{Arrington:2021vuu}. The number in parentheses is the normalization uncertainty, the solid curve is the parameterization of Ref.~\cite{Gomez:1993ri}, and the red band indicates the correlated systematic uncertainty for Ref.~\cite{Arrington:2021vuu}.} 
\label{fig:EMC-ratios-old}
\end{center}
\end{figure}

\begin{figure}[htb]
\begin{center}
\includegraphics[width=0.95\columnwidth,trim={0mm 4mm 0mm 2mm},clip]{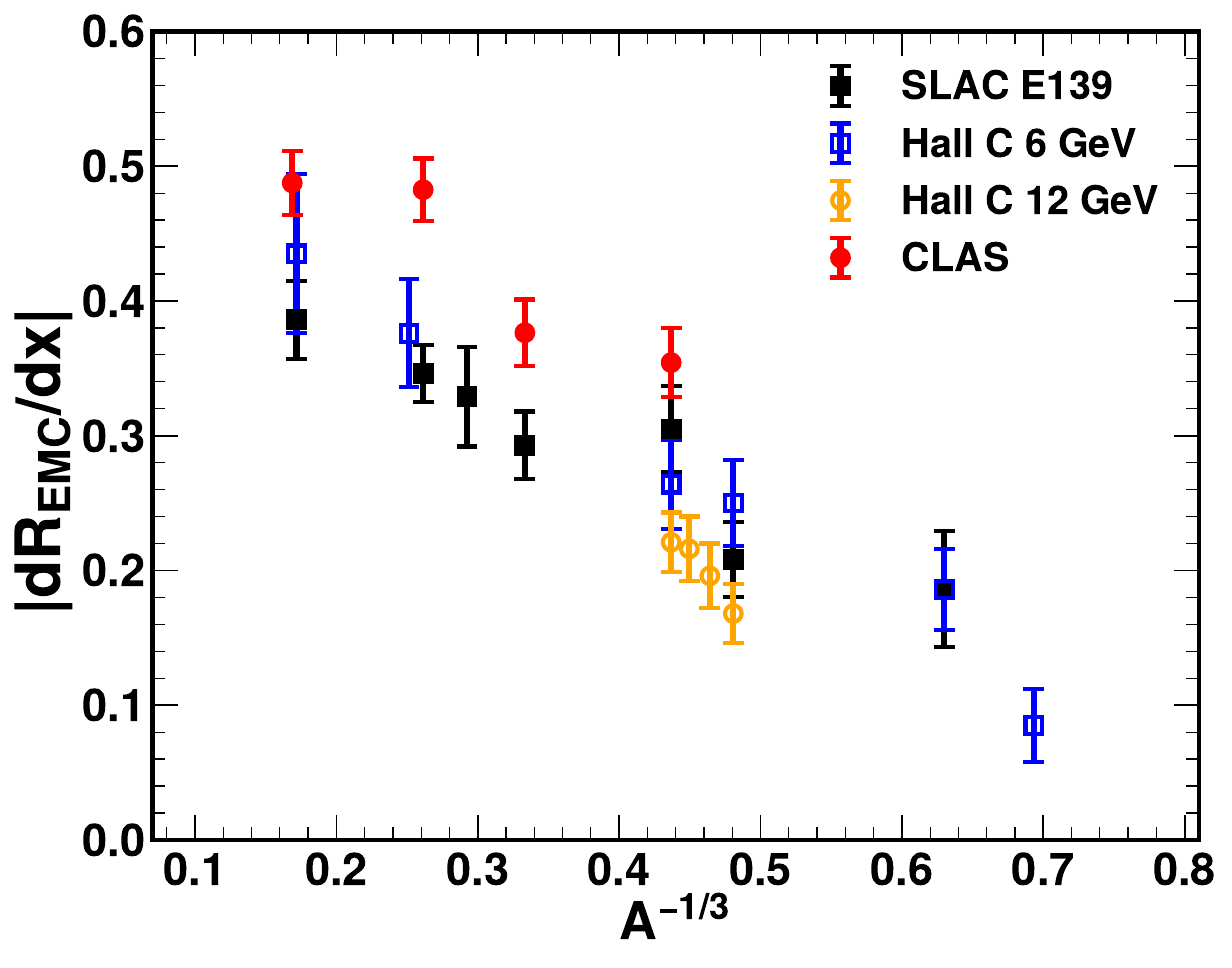}
\caption{EMC slopes from world's data, with the common isoscalar correction from Ref.~\cite{Arrington:2021vuu} applied to all data sets.}
\label{fig:EMC-slopes-old}
\end{center}
\end{figure}
 
Figure~\ref{fig:EMC-ratios-old} shows the EMC ratios from various experiments for heavy (A$\approx$200) nuclei after applying consistent isoscalar corrections~\cite{Arrington:2021vuu}. The size of the EMC effect is typically parameterized by taking the slope of the EMC ratios for $0.3\le x \le0.7$, and a comparison of world data on the EMC slopes is shown in Figure~\ref{fig:EMC-slopes-old}. As noted in Ref.~\cite{Arrington:2021vuu} the CLAS slopes~\cite{CLAS:2019vsb} are systematically higher, even when applying identical isoscalar corrections. Some fraction of the experimental uncertainty will be correlated across the slopes from a given measurement, as they all use common deuterium data. However, the CLAS slopes are systematically higher by roughly 0.10, much larger than the total quoted uncertainties of $\sim$0.02. 

Because the CLAS measurement~\cite{CLAS:2019vsb} includes only heavier nuclei, the inclusion of this data yields a significant change in the required A dependence over the full range of nuclei measured~\cite{Arrington:2021vuu}. Even excluding $^3$He, the inclusion of the CLAS data yields a significant difference in the A dependence. 
A linear fit of the EMC slope vs A$^{-1/3}$ for the data shown in Fig.~\ref{fig:EMC-slopes-old} (excluding $^3$He and $^4$He) yields a large reduced chi-squared value, $\chi^2_\nu=4.06$  when including the CLAS data, and $\chi^2_\nu=1.07$ without the CLAS points. Inclusion of the CLAS results increases the slope parameter (A dependence) of the fit by 22\%, a 2$\sigma$ increase, and the extrapolation to A~$=\infty$, corresponding to the EMC slope for infinite nuclear matter, is 15\% higher (2.8$\sigma$) than without the CLAS data. The impact of the CLAS data on both the result and quality of the fit demonstrates the importance of understanding potential differences between the various experiments. In this work, we examine the world's data on the EMC effect at large $x$ to try to understand these inconsistencies, as such differences can have a significant impact on understanding the A dependence of the EMC effect.

There are several differences between CLAS and the other experiments. CLAS has a lower beam energy and lower $Q^2$ values, a somewhat different $x$ range, and uses a large-acceptance detector as opposed to small-acceptance spectrometers. In addition, the experiments used somewhat different radiative corrections in extracting the cross-section ratios~\cite{Arrington:2021vuu}.

Some of these are straightforward to rule out as a significant contribution to the discrepancy. The CLAS data have a greater slope even when comparing over $0.3 \le x \le 0.6$, where all measurements have data, as seen in Fig.~\ref{fig:EMC-ratios-old}. While published results from different experiments applied different isoscalar corrections and Coulomb corrections, the cross-section ratios and slopes in Figs.~\ref{fig:EMC-ratios-old} and~\ref{fig:EMC-slopes-old} include identical isoscalar corrections~\cite{Arrington:2021vuu}, and the CLAS and JLab data used identical Coulomb corrections with the same corrections applied to the SLAC data in Ref.~\cite{Arrington:2021vuu}. Ref.~\cite{Arrington:2021vuu} also examined other effects that will be different at the lower $Q^2$ values of the CLAS measurement, e.g. quasielastic (QE) contribution to the EMC region and target-mass corrections, and showed that these do not explain a significant part of the observed difference. While the acceptance corrections are large, they have significant cancellations in the ratio.  So for the EMC ratios, the systematic uncertainties of the measurement~\cite{CLAS:2019vsb} are small enough that they cannot explain a large part of the discrepancy, even if the $x$ dependence is assumed to maximize the impact on the slopes. 

\section{Radiative Corrections}

Having argued that several differences are unlikely to explain the discrepancy, we focus on the difference in the radiative corrections (RCs). These are a natural candidate as the RCs introduce a target-dependent correction to the $x$-dependence of the EMC ratios. Because the EMC effect relies on precision comparisons of the $x$ dependence for different nuclei, it is sensitive to any correction that is $x$ and A dependent. Radiative corrections can have a significant $x$ dependence and are sensitive to the radiation length of the target. Several things about the CLAS radiative correction procedure are different from the other experiments. In addition to using a different RC formalism, the corrections will be different due to the lower beam energy of the CLAS measurement and the large acceptance (yielding a large $Q^2$ range contributing to each $x$ bin). In addition, the experiment used a dual-target system, where the liquid deuterium (LD2) target is in the beam upstream of one of the solid targets, leading to modified radiative corrections due to the LD2 target acting as an additional radiator for the solid target. 

The JLab Hall A and C measurements~\cite{Seely:2009gt, Arrington:2021vuu, HallC:2023xkj, JeffersonLabHallATritium:2021usd} use the RC code ``EXTERNALS'', which is based on the Mo and Tsai formalism~\cite{Tsai:1971qi} with details of the implementation described in Ref.~\cite{Dasu:1993vk}. This is essentially the same code that was used for radiative corrections for the SLAC E139~\cite{Gomez:1993ri} and E140~\cite{Dasu:1993vk} experiments, with only minor modifications. The CLAS analysis~\cite{CLAS:2019vsb} also used RCs based on the Mo and Tsai formalism, but these were implemented in the code ``INCLUSIVE''~\cite{inclusive}) that makes different approximations. 

The primary difference between the EXTERNALS and INCLUSIVE RC evaluation is how they evaluate the corrections over the full phase space that can contribute to a given event.  Events can radiate in from a 2D region in $(E_{beam},E_{e'})$ space, where the initial beam energy ($E_{beam}$) is higher than the energy at the scattering vertex and/or the scattered electron energy at the vertex is higher than observed at the detector ($E_{e'}$) due to the emission of real photons.  EXTERNALS integrates the contribution over the full 2D region, while INCLUSIVE uses the so-called ``energy-peaking approximation'' that replaces the 2D integral with a pair of 1D integrals. Further details and a more technical comparison of the two approaches are provided in the supplementary material~\cite{supplemental}.

In addition to using a different prescription for the radiative correction calculation, INCLUSIVE does not allow for materials other than the target to contribute to the external Bremsstrahlung.  This is particularly relevant for the CLAS measurements since the experiment used a target system that allowed the insertion of the liquid deuterium target and various solid targets in the beam at the same time~\cite{Hakobyan:2008zz}.  When calculating the radiative corrections for the solid targets, the INCLUSIVE program does not include the effects of external radiation in the upstream deuterium target.

We examined the impact of the RC calculation by comparing the INCLUSIVE calculation, as applied in Ref.~\cite{CLAS:2019vsb}, to the result from the EXTERNALS code used in the analysis of the other JLab extractions~\cite{Seely:2009gt, Arrington:2021vuu, HallC:2023xkj}. As a first step, we calculated the RC for the CLAS measurement using the INCLUSIVE code, with all parameters and settings matching those of the original publication, and verified that we obtained consistent results. This tests that our procedure for taking the cross-section weighted average RC as a function of $x$, integrated over the CLAS $Q^2$ acceptance, reproduces the original calculation. We then compare our INCLUSIVE RC factors to those obtained using EXTERNALS, with the inclusion of the liquid deuterium (LD2) target upstream of the solid target.

\begin{figure}[htb]
\begin{center}
\includegraphics[width=0.93\columnwidth,trim={0mm 4mm 0mm 2mm},clip]{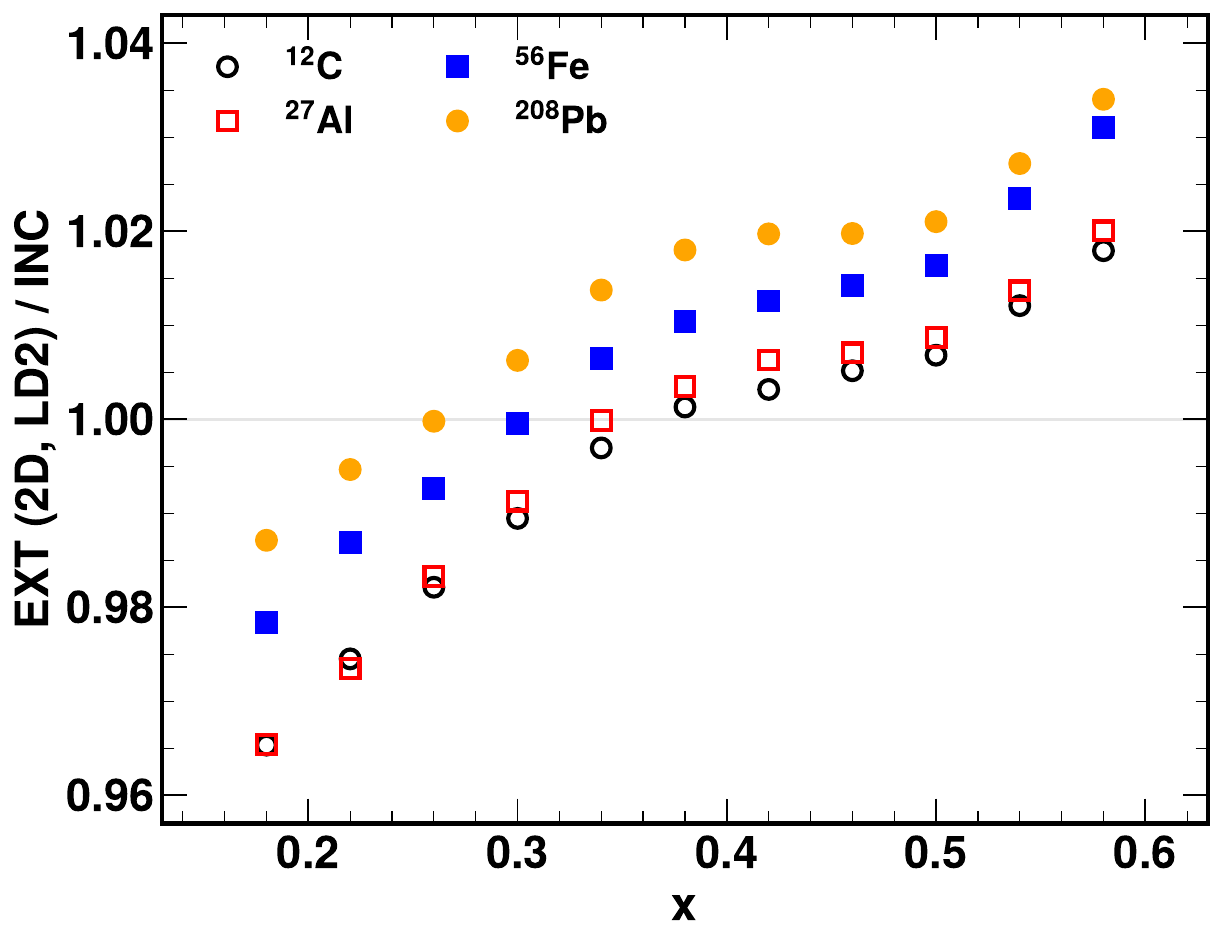}
\caption{Impact of the RC procedure (EXTERNALS, including upstream LD2 target) vs original (INCLUSIVE, no LD2 target) on the EMC ratios.}
\label{fig:RC-comparison-full}
\end{center}
\end{figure}

Figure~\ref{fig:RC-comparison-full} shows the change to the RC factors for the EMC ratios when applying the calculation of EXTERNALS (including the upstream LD2 target), as opposed to the INCLUSIVE corrections~\cite{CLAS:2019vsb}. The C and Al (hollow symbols) targets have thicknesses below 1\% of a radiation length (RL), while the Fe and Pb targets (solid symbols) have thicknesses of $\approx$2\% RL. The modification to the EMC ratio is typically $\ltorder$2\% and has a systematic $x$ dependence that is similar for all of the targets, decreasing the extracted EMC slopes.

The largest correction comes from the difference between the EXTERNALS and INCLUSIVE RC procedures. This introduces a 3\% $x$-dependence, roughly linear in $x$. It also introduces a small overall offset between low- and high-radiation length targets. The inclusion of the LD2 target as an upstream radiator yields a nearly identical correction for all targets, as expected, with a minimal impact on the RCs at larger $x$ and a reduction of 2\% at $x\approx0.2$. This yields a common distortion in the shape below $x=0.4$ for all targets but has only a small reduction in the EMC slopes when fitting over $0.3 \le x \le 0.7$. The supplemental material~\cite{supplemental} shows separately the impact of changing to the EXTERNALS correction code and the impact of including the LD2 target upstream. 

\section{Impact on the EMC ratios}

\begin{figure}[htb]
\begin{center}
\includegraphics[width=0.95\columnwidth,height=6.1cm,trim={3mm 0mm 16mm 18mm}, clip]{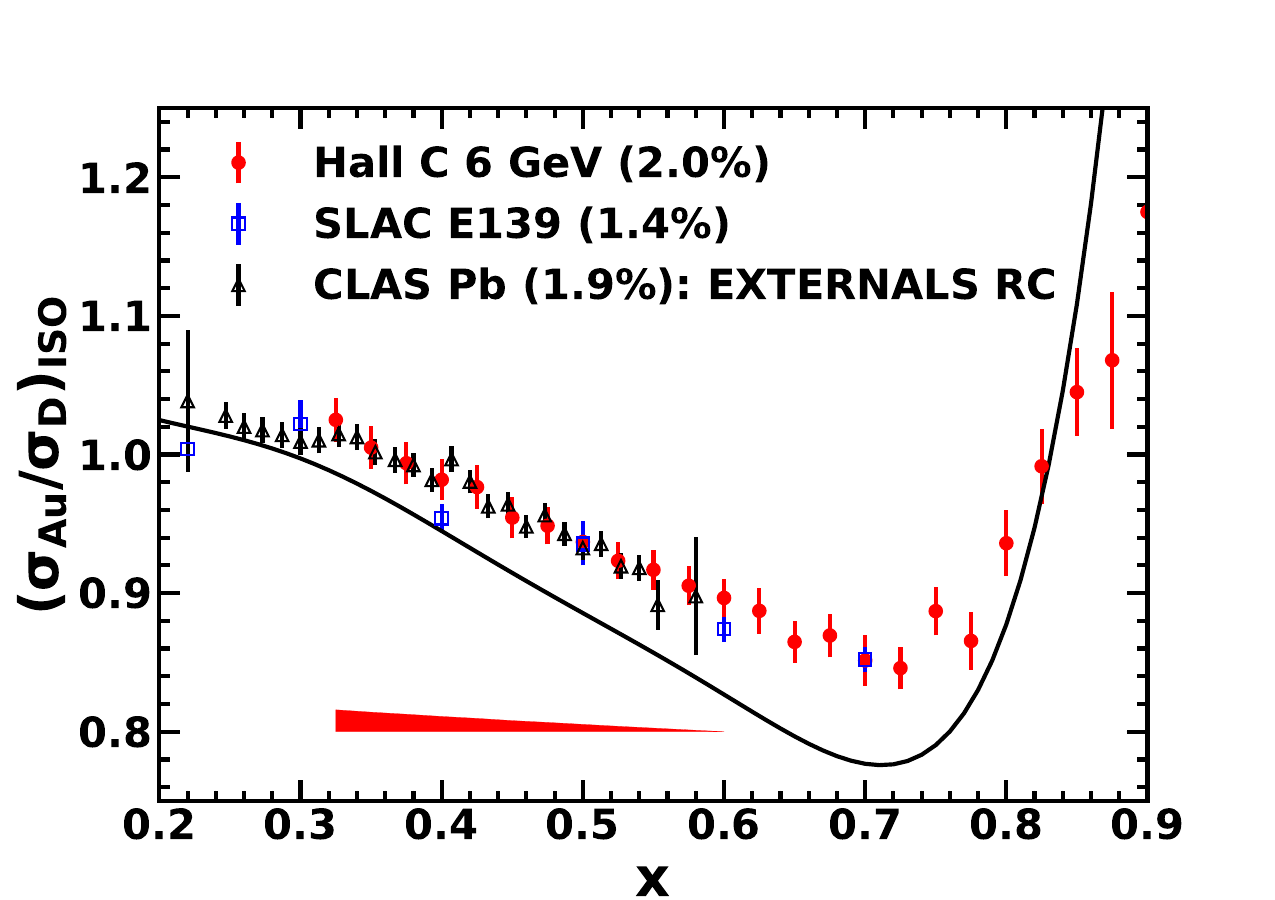}
\caption{EMC ratios for gold and lead~\cite{Gomez:1993ri, CLAS:2019vsb, Arrington:2021vuu} with common isoscalar corrections from~\cite{Arrington:2021vuu} and the radiative correction procedure described in the text applied to the CLAS data. Symbols as in Fig.~\ref{fig:EMC-ratios-old}.}
\label{fig:EMC-ratios-new}
\end{center}
\end{figure}

\begin{figure}[htb]
\begin{center}
\includegraphics[width=0.95\columnwidth,trim={0mm 4mm 0mm 2mm}, clip]{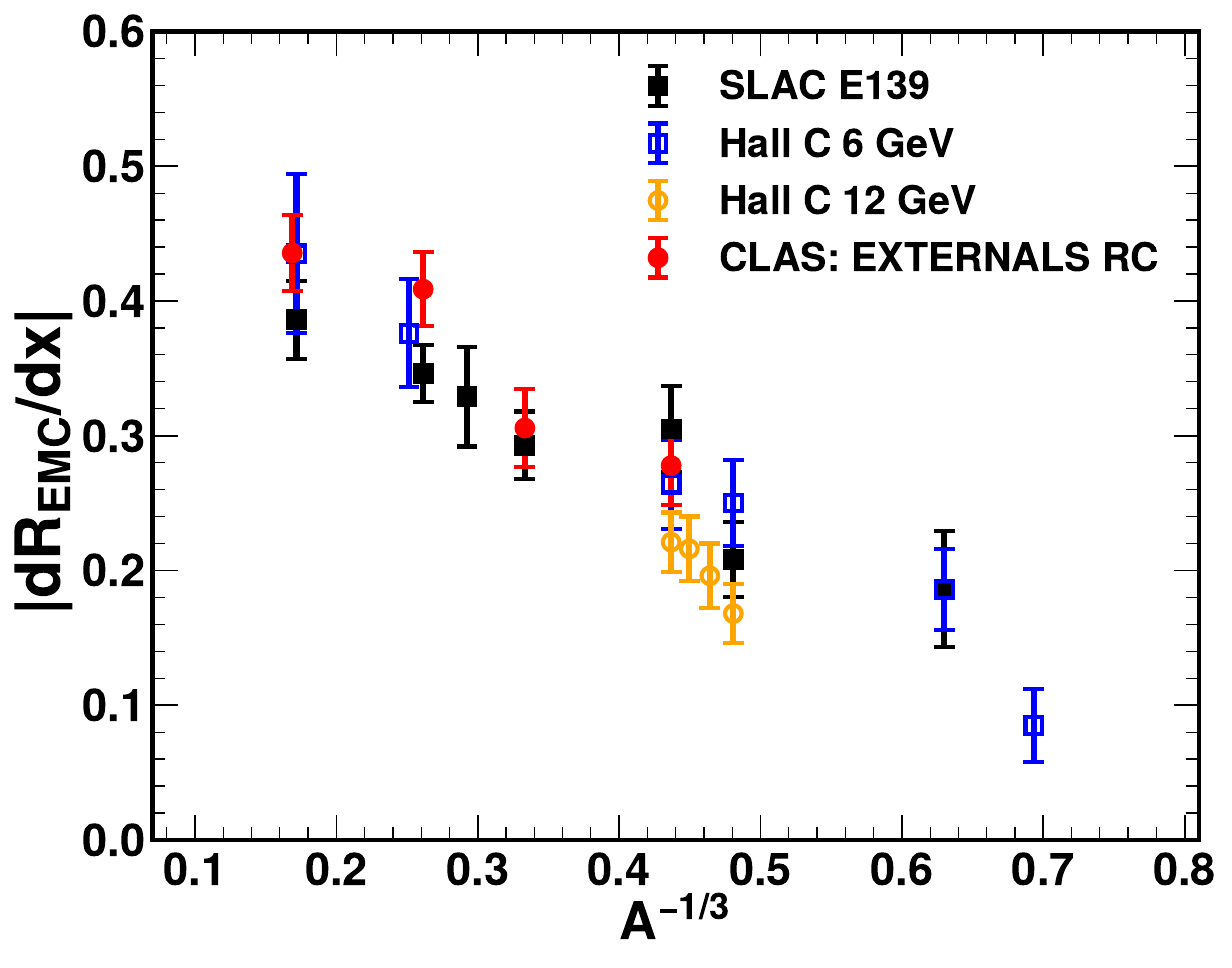}
\caption{EMC slopes from world's data, with the common isoscalar correction from Ref.~\cite{Arrington:2021vuu} and the radiative correction procedure described in the text applied to the CLAS data. Symbols as in Fig.~\ref{fig:EMC-slopes-old}.}
\label{fig:EMC-slopes-new}
\end{center}
\end{figure}

We note that there was also a cryogenic isolation foil between the LD2 target and the solid targets which we do not include in this study, but its radiation length was very small compared to the LD2 target and the upstream material had minimal impact except at very low $x$~\cite{supplemental}, so this has a negligible impact on our study. In addition, the upstream LD2 target was empty for the Aluminum data taking, reducing its radiation length and thus the impact on the radiative corrections. So the correction for the Aluminum target slightly overestimates the effect, but again the upstream radiator has minimal impact on the EMC slope.

Given that the EXTERNALS calculation yields a different correction factor for the CLAS data, we now evaluate the impact of these changes on the extracted EMC ratios and slopes. To ensure that all other aspects of the analysis are unchanged, we start with the published cross-section ratios from~\cite{CLAS:2019vsb}, which used the INCLUSIVE RCs.  We then divide out our INCLUSIVE RC factor and multiply by the EXTERNALS RC factor to provide EMC ratios based on the EXTERNALS correction. To ensure a consistent comparison with other measurements, we follow the approach of Ref.~\cite{Arrington:2021vuu} and ensure that all data sets use the same correction.

While the uncertainty in the $x$-dependent radiative correction will partially cancel between different targets, this cancellation will not be perfect. We add 0.5\% uncorrelated uncertainty to each point to account for uncertainties in our procedure that could introduce a potential bias in the $x$ dependence of the ratios. The sensitivity to the quasielastic model is somewhat smaller than that seen for the Hall C measurement (red band in Fig.~\ref{fig:EMC-ratios-old}), due to the lower $Q^2$ of the CLAS measurements and lower radiation length targets.

Figure~\ref{fig:EMC-ratios-new} shows the cross-section ratios for worlds data on Au and Pb, using the ratios from Ref.~\cite{Arrington:2021vuu} and applying the EXTERNALS RC to the CLAS data. We added in quadrature our estimated uncertainty for the radiative correction procedure. Figure~\ref{fig:EMC-slopes-new} shows the EMC slopes from fitting the EXTERNALS-corrected CLAS EMC ratios for $x \ge 0.3$. The EXTERNALS-corrected result yields better agreement in the EMC ratios and the EMC slope for all nuclei measured. The slopes are reduced by a significant amount, as seen in the comparison to Fig.~\ref{fig:EMC-slopes-old}, and are now in good agreement with the remaining world's data.  The supplemental material~\cite{supplemental} includes the data shown in Figs.~\ref{fig:RC-comparison-full}-\ref{fig:EMC-slopes-new}.

Repeating the linear fit of EMC slope vs A$^{-1/3}$ for A$\ge$9 nuclei using the EXTERNALS-corrected CLAS slopes yields $\chi^2_\nu=1.22$, only slightly above the value for the fit excluding CLAS, and the slope and intercept of the fit are within 1$\sigma$ of the values for the fit without the CLAS results. This gives an EMC slope for A~$=\infty$ of 0.542$\pm$0.023.

\section{Radiative corrections for quasielastic scattering at $x>1$}

Having observed that the RC procedure has a significant impact on the EMC ratios and their $x$ dependence, we performed a similar comparison of the RCs in the region where scattering from short-range correlations (SRCs) are expected to dominate~\cite{frankfurt88, sargsian03, Arrington:2012ax, fomin17, Arrington:2022sov}. Ref.~\cite{CLAS:2019vsb} extracted these ratios in addition to the DIS ratios at lower $x$, and so the same differences that changed the EMC ratios could also modify the ratios in the SRC region. We repeat the comparisons presented above for the data in the SRC region to determine if it leads to similar systematic differences in the comparison to previous SRC measurements~\cite{frankfurt93, Fomin:2011ng, Li:2022fhh}. In this region, the goal is to examine the ratio of the cross-section from heavy nuclei to the deuteron, which is expected to be constant for large $x$ and $Q^2$, where scattering is dominated from quasielastic-scattering off of high-momentum nucleons from SRCs~\cite{frankfurt93,Arrington:2022sov}. 

In this case, there is less reason to expect that the impact will be large. While the inclusion of the LD2 target upstream could produce a systematic shift in the ratios, the uncertainties in the average cross-section ratios in the SRC region, $a_2$(A), are roughly 5\%. So the 2\% level changes observed in the DIS region would have a much less significant impact on the relative precision of the SRC measurements. In addition, because the radiative corrections at $x>1$ are dominated by the loss of high-$x$ events that radiate into lower-$x$ kinematics, with very little radiating in from higher $x$ where the cross sections are highly suppressed, there is less sensitivity to the cross-section model. This would tend to reduce the A dependence but not the potential dependence on the target thickness. Figures showing the impact on the cross-section ratios at $x>1$ and modified versions of the extracted $a_2$(A) ratios are included in the supplemental material~\cite{supplemental}. The main impact is a systematic decrease in the extracted $a_2$(A) values of up to 1-2\%, compared to the $\sim$2-3\% uncertainty in the original extraction. Thus, it is a small effect, although not negligible given that it is partially correlated between the different nuclei, as shown in the supplemental material~\cite{supplemental}.

\section{Conclusions}

In conclusion, we have performed comparisons of the INCLUSIVE and EXTERNALS radiative correction procedures, and find that the use of the more complete EXTERNALS formalism on the CLAS data~\cite{CLAS:2019vsb} appears to resolve the discrepancies between the EMC ratios from this measurement and those reported by SLAC and Jefferson Lab Hall C experiments. Focusing on the CLAS~\cite{CLAS:2019vsb} experiment, we find that employing an improved numerical-integration implementation of the Mo and Tsai radiative correction procedures, as done in EXTERNALS, yields corrections of up to 2-3\% with a systematic $x$ dependence. Additionally, the inclusion of the upstream CLAS LD2 target introduces a distortion in the shape at low $x$. While these corrections are relatively small, they are significant given the precision of the data~\cite{CLAS:2019vsb}. Because of the nearly linear $x$ dependence, these effects combine to give a systematic reduction in the slope of the EMC effect for all targets. The adoption of this alternative radiative correction procedure aligns the CLAS EMC slopes with earlier SLAC and Hall-C measurements. This yields more consistent results for individual targets, as well as making global fits to the A dependence of the EMC effect less sensitive to whether or not the CLAS results are included.  

These studies can also inform future measurements of hadron production in nuclear DIS such as Refs.~\cite{CLAS:2021jhm, CLAS:2022oux, CLAS:2022oux, HERMES:2000ytc, HERMES:2011qjb}, which typically feature observables that are normalized relative to the inclusive DIS cross-sections. They may also be of importance for other experiments that collect data with multiple targets in the beam. 

\section*{Acknowledgements}

This material is based upon work supported by the U.S. Department of Energy, Office of Science, Office of
Nuclear Physics under contracts DE-AC05-06OR23177, DE-AC02-05CH11231, and DE-SC0022324.

\bibliographystyle{apsrev4-1}
\bibliography{EMC-RC}

\clearpage

\section{Supplemental Material}

\subsection{Differences between the INCLUSIVE and EXTERNALS radiative corrections prescriptions}

\begin{figure}[htb]
\begin{center}
\includegraphics[width=0.9\columnwidth,height=5cm,trim={0 8mm 0 8mm},clip]{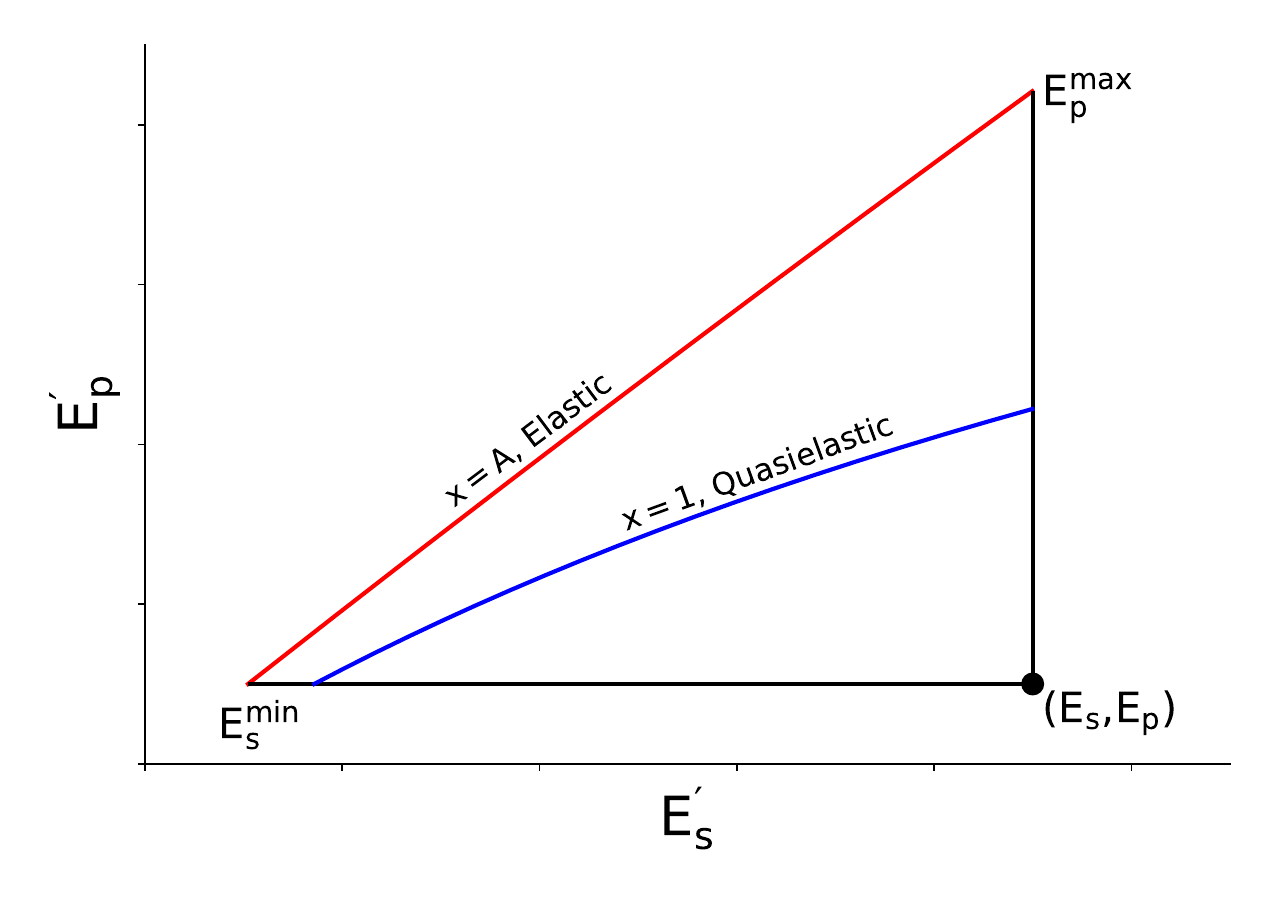}
\caption{Radiative corrections triangle. This diagram shows the phase space of initial and scattered electrons (after radiation) that can contribute to the radiated cross-section at $(E_{beam},E_{e})=(E_s,E_p)$.  The red line denotes the elastic limit where $x=A$ and the blue line is the location of the quasielastic contribution. In principle, nuclear-excited states can populate the region between the elastic and quasielastic, while the inelastic contributes to the region between the quasielastic and the measured point. In the energy-peaking approximation, the full two-dimensional integral over this phase space is simplified to two one-dimensional integrals along the $E_s^\prime$ and $E_p^\prime$ axes.}
\label{fig:RC_triangle}
\end{center}
\end{figure}

The primary difference between the EXTERNALS and INCLUSIVE programs is how each treats the so-called radiative correction ``triangle''.  The phase space of initial and scattered electron energies that can contribute to the cross-section at $(E_{beam},E_{e})=(E_s,E_p)$ after radiation is illustrated in Figure~\ref{fig:RC_triangle}.  Any point encompassed by the red curve and black lines in the figure (which includes nuclear elastic, quasielastic, and inelastic processes) can contribute to the measured cross-section. So a complete calculation of the radiated cross-section at $(E_s, E_p)$ requires a two-dimensional integral of the appropriately weighted cross-section over the phase space. The EXTERNALS program performs this two-dimensional integral 
while the INCLUSIVE program makes use of the so-called ``energy-peaking approximation'', which allows the simplification of the full two-dimensional integral to a pair of one-dimensional integrals along the $E_s$ and $E_p$ contours. This approximation is useful in cases in which there is insufficient existing data to constrain the cross-section away from the point of the measurement. However, the ``energy-peaking approximation'' is inadequate in certain cases, in particular for large radiation-length targets.

As described in Ref.~\cite{Tsai:1971qi}, the measured radiated cross-section from a target of thickness $T$, at a beam energy $E_b=E_s$ and scattered electron energy $E_e^\prime = E_p$ is given by,
\begin{multline}\label{rc_eqn}
    \sigma_{exp}(E_s,E_p) = \int_0^T \frac{dt}{T} \int_{E_s^{min}}^{E_s} dE^\prime_s \int_{E_p}^{E_p^{max}} dE^\prime_p \\ I(E_s,E_s^\prime,t) \sigma_I(E_s^\prime E_p^\prime) I(E_p^\prime,E_p,T-t),
\end{multline}
where $\sigma_I(E_s^\prime E_p^\prime)$ is the cross section including internal radiative corrections and $I(E,E-\delta,t)$ is the probability of an electron with energy $E$ to lose an amount of energy $\delta$ in material of thickness $t$ due to Bremsstrahlung and ionization energy loss.  The limits of integration are given by,
\begin{equation}
E_p^{max}(E_s^\prime) = \frac{E_s^\prime}{1+\frac{E_s^\prime}{M} (1-cos{\theta})},
\end{equation}
and
\begin{equation}
E_s^{min}(E_p) = \frac{E_p}{1-\frac{E_p}{M} (1-cos{\theta})},
\end{equation}
where $M$ is the target mass and $\theta$ is the electron scattering angle.  

The program EXTERNALS calculates the two-dimensional integral in Eq.~\ref{rc_eqn} over the full phase space, calculating $\sigma_I$ (the cross section including internal radiative corrections) using the so-called equivalent radiator approximation.

The energy-peaking approximation employed by INCLUSIVE reduces the two-dimensional integration to line integrals along the $E_p$ and $E_s$ axes. 
In some cases, this approximation does not result in large changes to the radiated cross section (on the order of 1\%).  For thick targets and certain kinematics, the effects can be significant, though.

\begin{figure}
    {\includegraphics*[width=0.4\textwidth, trim={9mm 9mm 6mm 8mm}, clip]{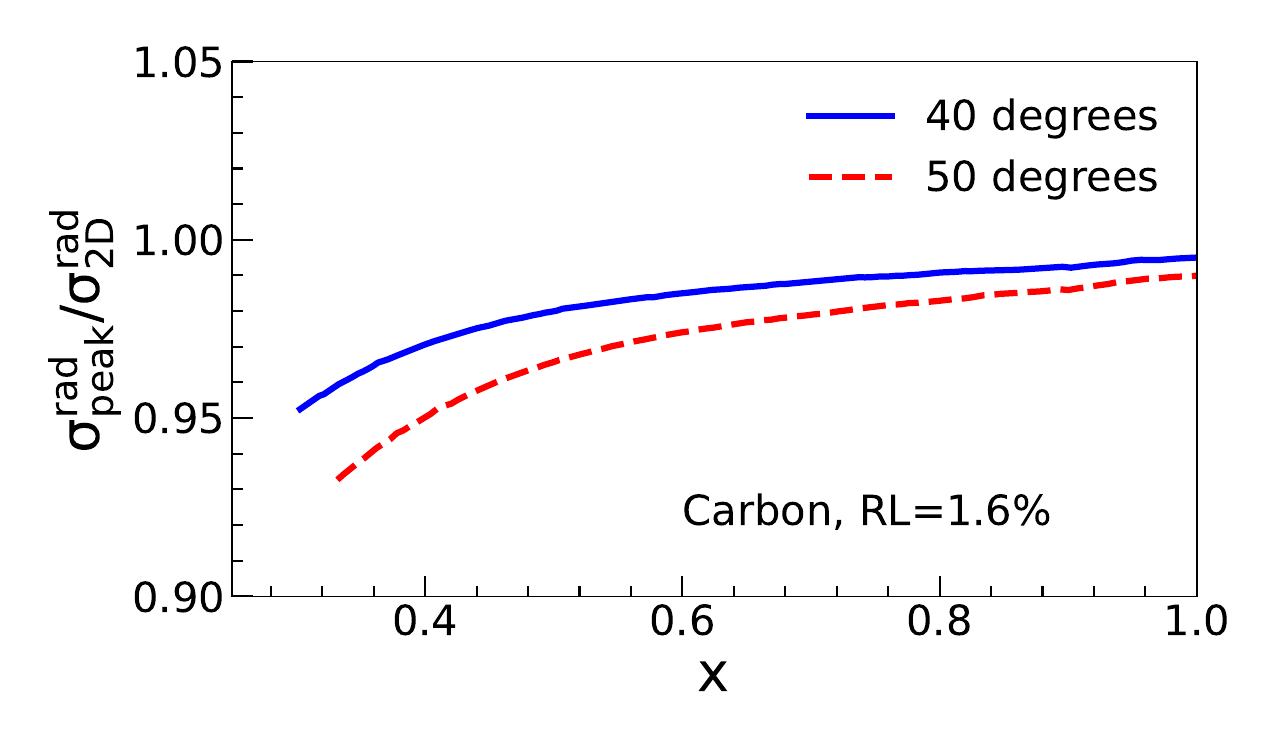}} \\
    {\includegraphics*[width=0.4\textwidth, trim={9mm 9mm 6mm 8mm}, clip]{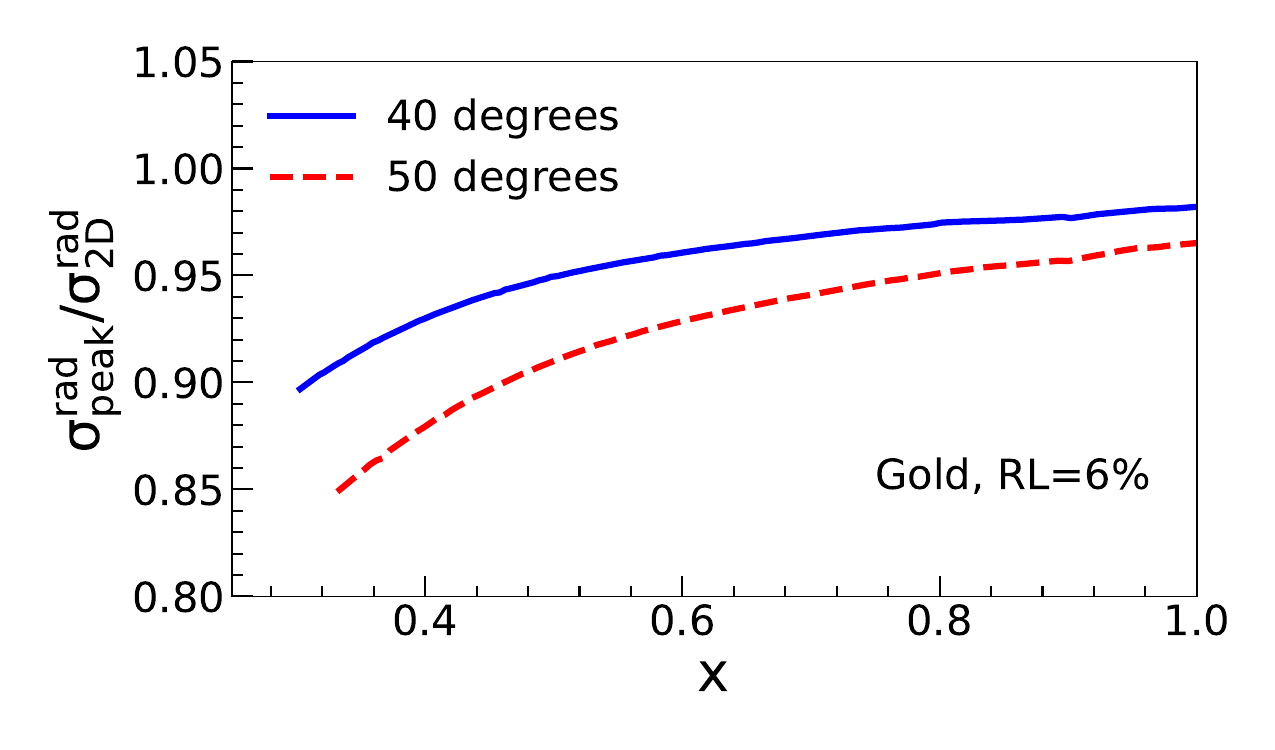}}
    \caption{Ratio of radiated cross section calculated using the energy-peaking approximation ($\sigma^{rad}_{peak}$) to that from the full 2-dimensional integral over the radiative corrections ``triangle'' ($\sigma^{rad}_{2D}$).  Calculations were performed at the kinematics of the Hall C 6 GeV EMC effect measurements (E03103) using the nominal carbon and gold target thicknesses (1.6\% and 6\% RL respectively).  Effects can be significant for certain kinematics (large angle and low $x$) and for thicker targets.}
    \label{fig:xem_rc_compare}
\end{figure}

A comparison of the radiated cross section calculated with and without the energy-peaking approximation is shown in Figure~\ref{fig:xem_rc_compare}. These calculations were done as part of the analysis of Hall C experiment E03103~\cite{Seely:2009gt, Arrington:2021vuu} at Jefferson Lab. In this experiment, measurements were made of the EMC effect at $E_{beam}=5.776$~GeV for several targets ($^3$He, $^4$He, $^9$Be, $^{12}$C, Cu, and $^{197}$Au). The $^9$Be and $^{12}$C targets had relatively small radiation lengths (1.6-3\%), while the Cu and $^{197}$Au targets were thicker (6\%). Fig.~\ref{fig:xem_rc_compare} shows the ratio of radiated cross sections calculated with and without the energy-peaking approximation for the $\approx$1.6\% RL carbon target (top) and the 6\% RL gold target (bottom).  The radiative effects shown in Fig.~\ref{fig:xem_rc_compare} were calculated using the same parametrization for the inelastic and quasielastic contributions~\cite{Bosted:2012qc}, the only difference being the application (or not) of the energy-peaking approximation. The impact of using the energy-peaking approximation is clearly larger for the larger RL target and at a larger scattering angle (larger $Q^2$). At large $x$ the difference is only a few percent, but at smaller $x$ ($\approx$0.3), the difference can grow to more than 10\%, so the energy-peaking approximation was not used in the E03103 analysis.  The kinematics of the CLAS measurement are similar to those from the Hall C measurement, suggesting that the use of the energy-peaking approximation might induce biases. 

\subsection{Detailed radiative correction comparisons}

\begin{figure}[htb]
\begin{center}
\includegraphics[width=0.8\columnwidth,trim={0mm 0mm 0mm 0mm}, clip]{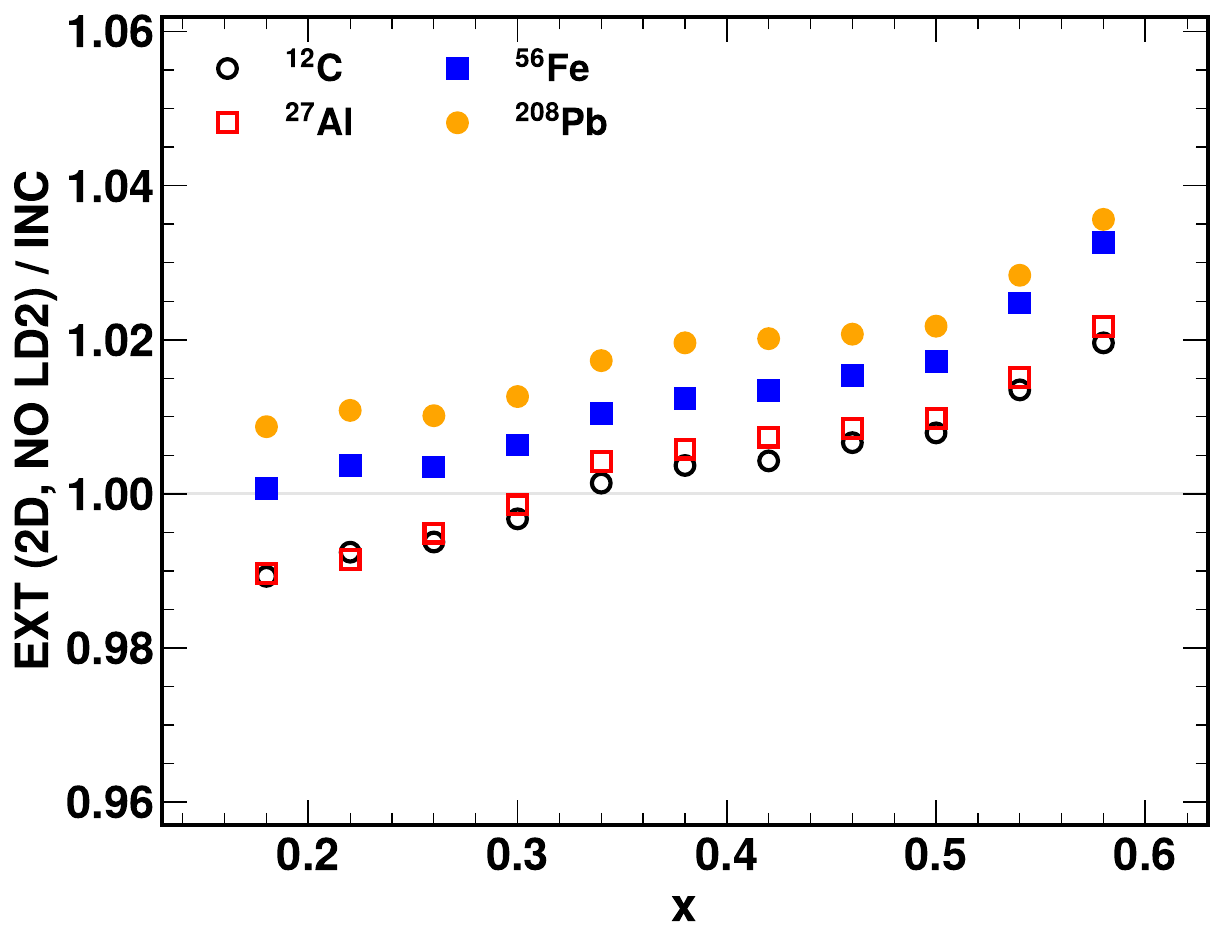}
\caption{Ratio of radiative correction factors for the targets used in the CLAS EMC effect extraction calculated using the EXTERNALS program (but without including the LD2 target that was upstream of the solid targets) relative to the RC factor calculated using INCLUSIVE (i.e., making use of the energy peaking approximation).}
\label{fig:RC-EXT-vs-INC}
\end{center}
\end{figure}

The main paper shows the impact of using the EXTERNALS radiative correction procedure and including the upstream LD2 target (Fig.~\ref{fig:RC-comparison-full}) on the EMC ratios. Here, we break down the different contributions to the modification of the radiative corrections. Figure~\ref{fig:RC-EXT-vs-INC} shows the impact of switching from the INCLUSIVE to EXTERNALS RC code, without the inclusion of the upstream LD2 target as an additional external radiator.  For all targets, this introduces a correction that is roughly linear in $x$ and decreases the falloff of the EMC ratio from $x=$0.3--0.7 by 2--3\%, thus reducing the EMC slope. The $x$ dependence of the modification to the EMC ratios is similar for all of the targets, although the thick targets have a 1--2\% change in the overall normalization of the correction. These changes bring the slope in better agreement with world's data, and also explain some of the normalization difference~\cite{Arrington:2021vuu}.

\begin{figure}[htb]
\begin{center}
\includegraphics[width=0.8\columnwidth,trim={0mm 0mm 0mm 0mm}, clip]{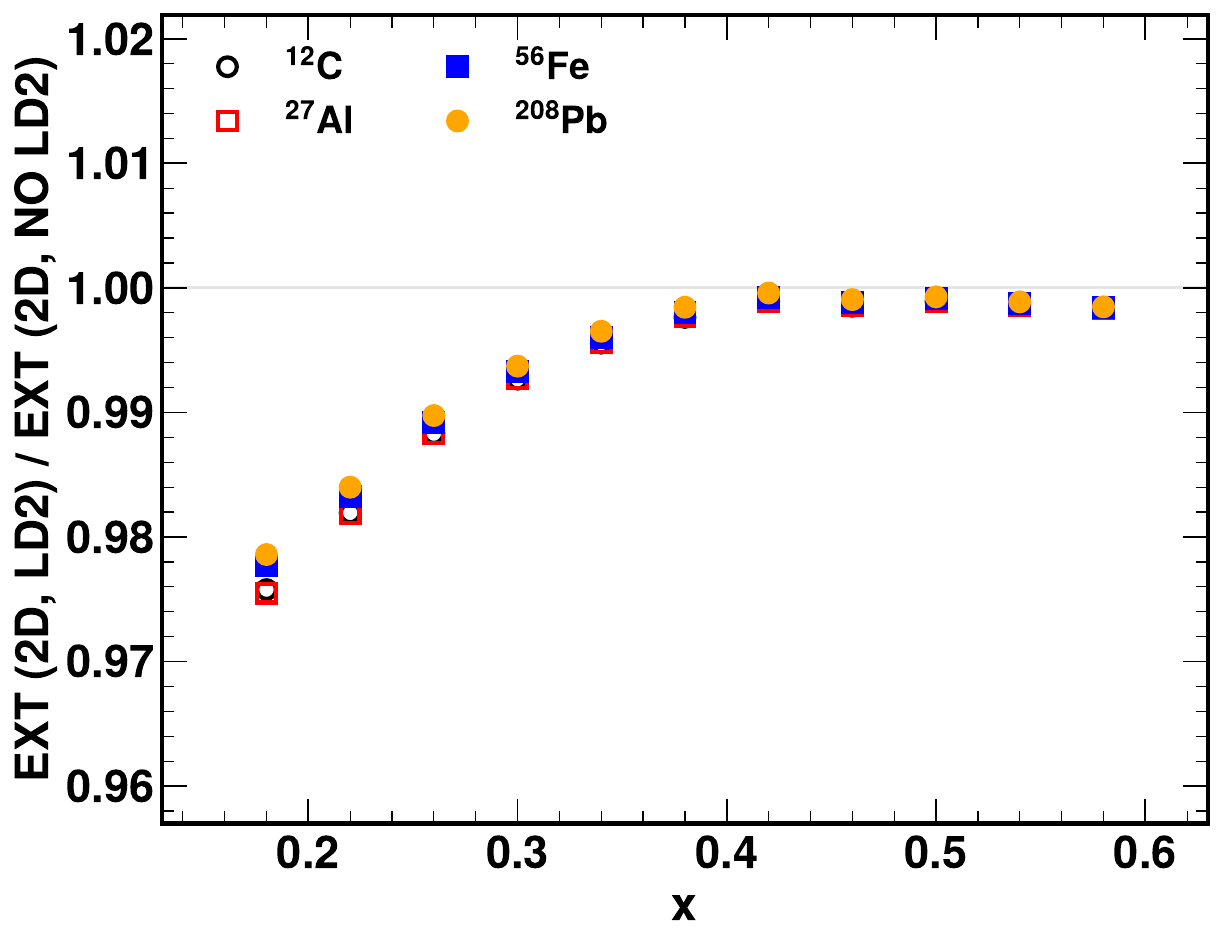}
\caption{Ratio of radiation correction factors calculated using EXTERNALS with and without the upstream LD2 target included in the calculation.}
\label{fig:EMC_1Panels_EXT_LD2_NoLD2_comp}
\end{center}
\end{figure}

Figure~\ref{fig:EMC_1Panels_EXT_LD2_NoLD2_comp} shows the impact of adding the LD2 target upstream of the solid target position, evaluated using the EXTERNALS code. The main impact is a reduction of the EMC ratio at low $x$ values, mainly below $x=0.4$. The correction is nearly A independent, and is generally a 1\% effect or less in the EMC region ($x \ge 0.3$), yielding a small reduction in the EMC slope that is largely target independent. If data below $x=0.3$ is used in fitting the EMC effect, or as a consistency check with other data sets, the impact would be larger - up to 2.5\% for the lowest $x$ values.

The total impact to the RC shown in Fig.~\ref{fig:RC-comparison-full} consists of combining the contributions shown in Figs.~\ref{fig:RC-EXT-vs-INC} and~\ref{fig:EMC_1Panels_EXT_LD2_NoLD2_comp}. This can then be applied as a multiplicative correction to the CLAS EMC ratios from Ref.~\cite{Arrington:2021vuu} which after applying common isoscalar corrections for all data sets. 

\subsection{Corrections for quasielastic scattering at $x>1$}

\begin{figure}[htb]
\begin{center}
\includegraphics[width=0.8\columnwidth,trim={0mm 0mm 0mm 0mm}, clip]{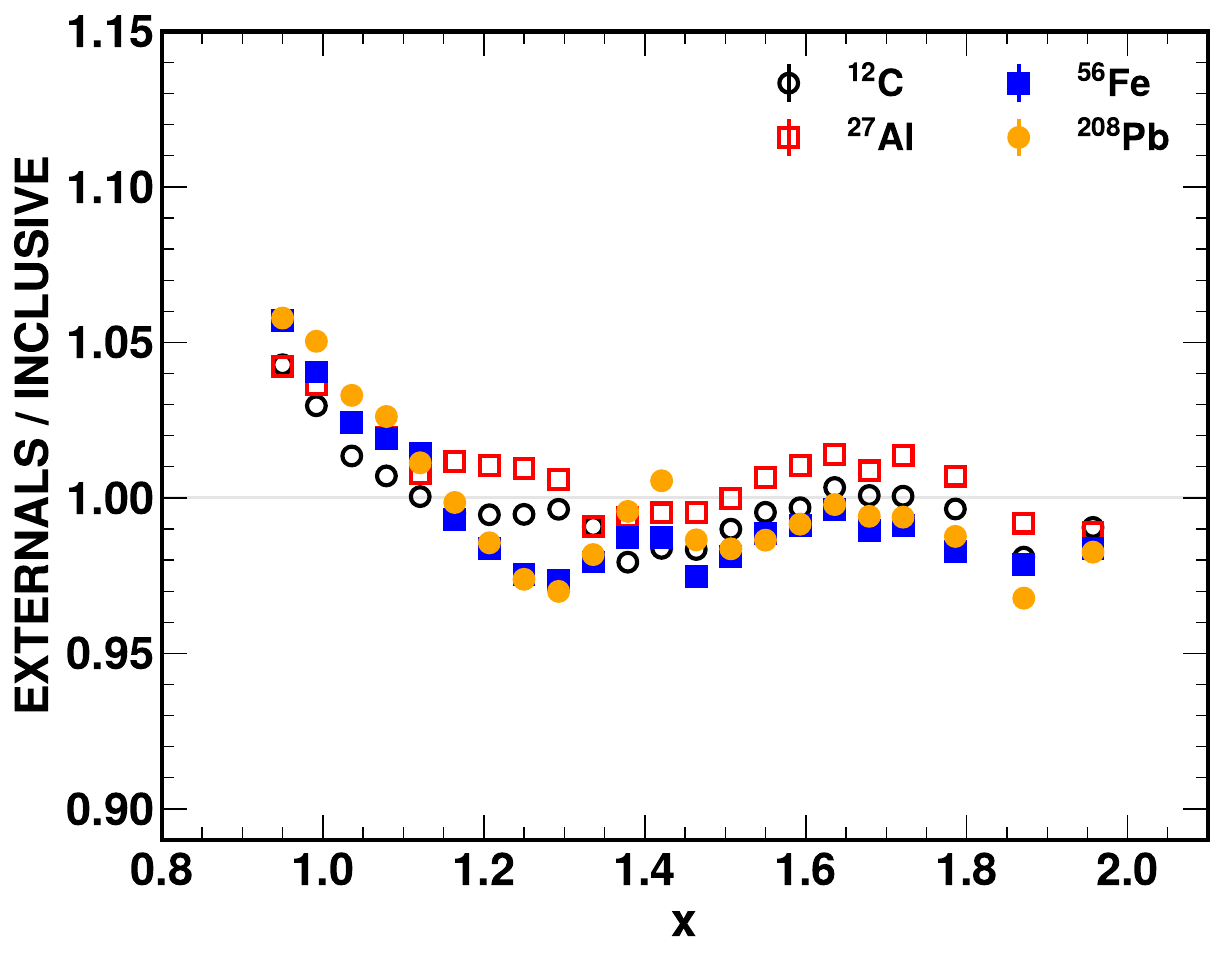}
\caption{Comparison of the EXTERNALS RC correction (including upstream LD2 target) vs original (INCLUSIVE, no LD2 target) in the SRC region.}
\label{fig:SRC-EXT-INC}
\end{center}
\end{figure}

\begin{figure}[htb]
\begin{center}
\includegraphics[width=0.8\columnwidth,trim={0mm 0mm 0mm 0mm}, clip]{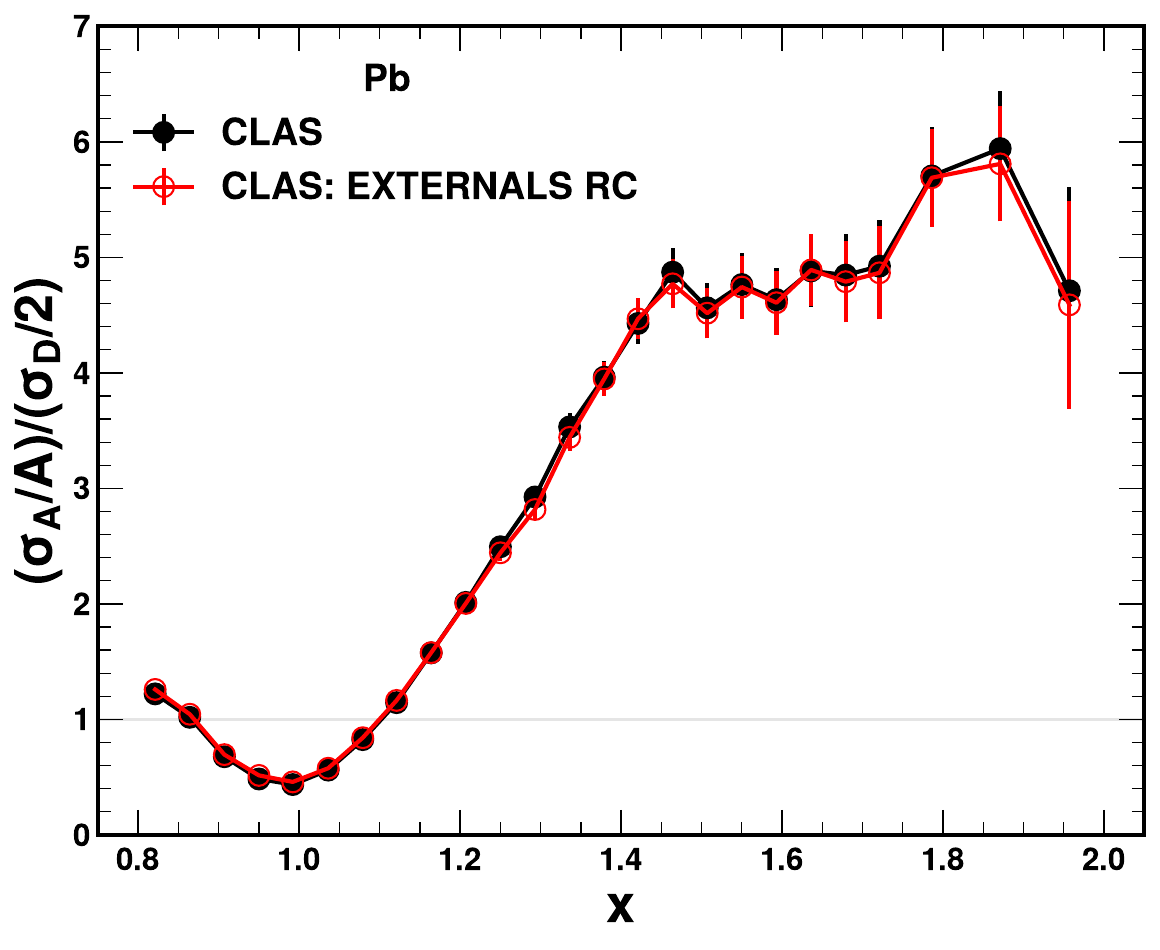}
\caption{Pb/D ratio for $x>1$ for the CLAS data~\cite{CLAS:2019vsb} using the INCLUSIVE (black) and EXTERNALS (red) radiative corrections. }
\label{fig:SRC-before-after}
\end{center}
\end{figure}

\begin{figure}[htb]
\begin{center}
\includegraphics[width=0.95\columnwidth,trim={0mm 0mm 0mm 0mm}, clip]{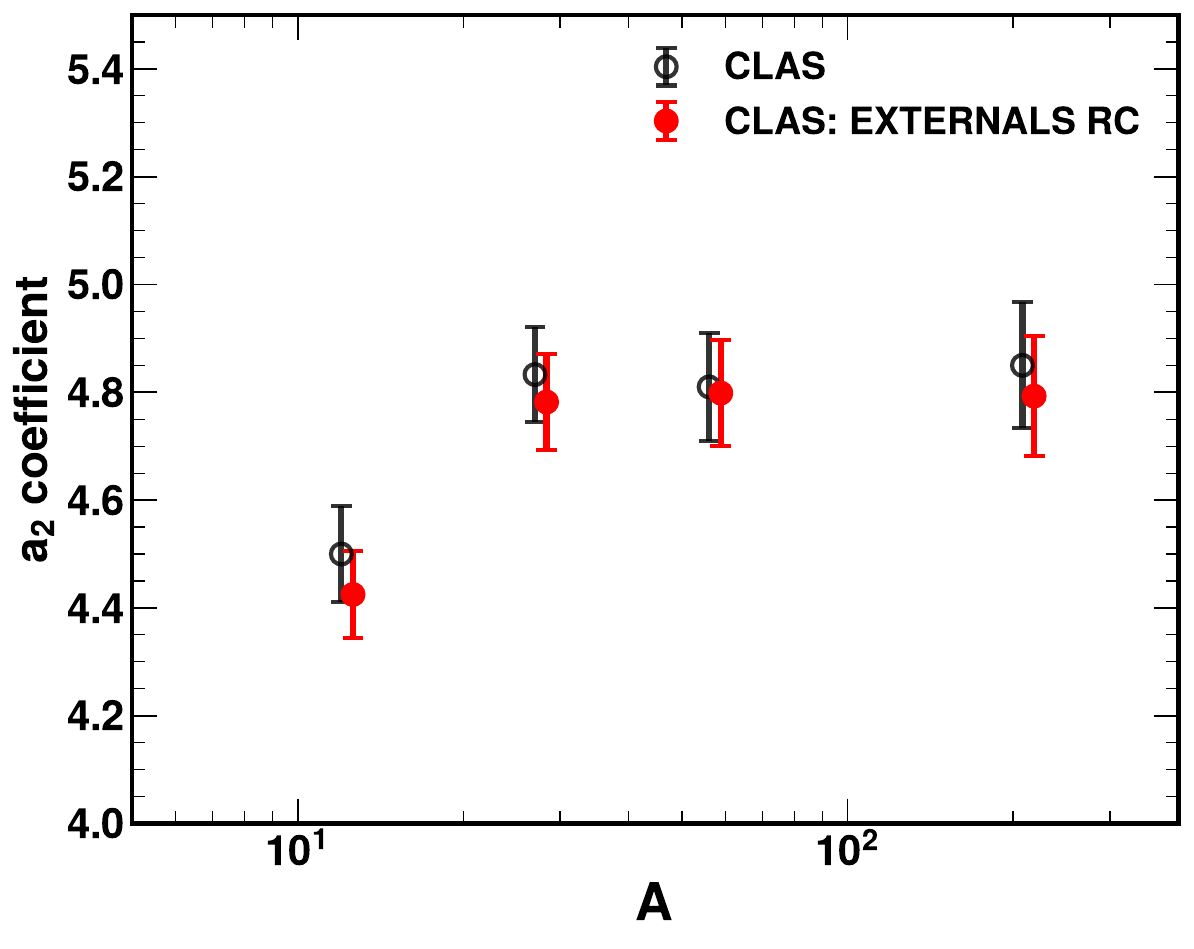}
\caption{Impact of RC procedures described in the text on SRC plateau values, $a_2(A)$. The $a_2$ values based on the EXTERNALS RC are offset slightly for clarity. }
\label{fig:SRC-a2}
\end{center}
\end{figure}

Next, we present a similar study for quasielastic scattering at $x>1$. In this case, the physics interpretation of the data in terms of short-range correlation does not require the same level of precision as the in the DIS region. Thus, even if the difference between EXTERNALS and INCLUSIVE could be as large or even larger, the impact is likely to be less.  Figure~\ref{fig:SRC-EXT-INC} shows the change to the A/D ratios when converting from INCLUSIVE to EXTERNALS, where the EXTERNALS code includes the addition of the upstream LD2 target. The correction is largest near the top of the QE peak, where the rapidly changing cross-section (and ratio) in the QE peak region makes the corrections more sensitive to the RC procedure.  The correction is small and relatively $Q^2$ independent above $x=1.3$, the region of interest for SRC studies. The overall impact in this region is a reduction of up to 2\% in the cross-section ratio that is somewhat larger for the heavy (higher radiation length) targets.

Figure~\ref{fig:SRC-before-after} shows the Pb/D ratio as published~\cite{CLAS:2019vsb} (``CLAS'') and after the application of the EXTERNALS radiative correction, and Figure~\ref{fig:SRC-a2} shows the impact on the extracted value of $a_2(A)$ for all four nuclei measured in the CLAS experiment.  While the RC changes led to a reduction of up to 2\% in the SRC-dominated region, this is not large compared to the experimental uncertainties on the $a_2$ extraction. Nonetheless, because it is a systematic reduction for all nuclei, with some A dependence, it is not completely negligible.

Table~\ref{tab:EMCslopes} give the EMC slopes extracted from the EXTERNALS-corrected CLAS ratios, as shown in Fig.~\ref{fig:EMC-slopes-new}. Table~\ref{tab:EMCratios} shows the ratio of EXTERNALS to INCLUSIVE radiative corrections from Fig.~\ref{fig:RC-EXT-vs-INC} and the EXTERNALS-corrected EMC ratios from Fig.~\ref{fig:EMC-ratios-new}.

\begin{table}[hb]
   \centering
    \begin{tabular}{|c|c|c|}
    \hline
    Target & Slope (EXTERNALS)  & $a_2$(A) (EXTERNALS)\\ 
    \hline
    $^{12}$C   & 0.278 $\pm$ 0.030  &  4.43 $\pm$ 0.081 \\
    $^{27}$Al  & 0.306 $\pm$ 0.029  &  4.78 $\pm$ 0.089 \\
    $^{56}$Fe  & 0.409 $\pm$ 0.027  &  4.80 $\pm$ 0.099 \\
  ~$^{208}$Pb~& 0.436 $\pm$ 0.028   &  4.79 $\pm$ 0.110 \\
    \hline
  \end{tabular}
  \caption{EMC slopes (absolute value) and $a_2$(A) for the EXTERNALS-corrected CLAS data, using the isoscalar corrections from Ref.~\cite{Arrington:2021vuu}}.
  \label{tab:EMCslopes}
\end{table}

\begin{table*}[htb]
    \centering
    \begin{tabular}{|c|c|cc|cc|cc|cc|}
    \hline
        $x$ & $\langle Q^2 \rangle$ & \multicolumn{2}{c|}{$^{12}$C/D} & \multicolumn{2}{c|}{$^{27}$Al/D} & \multicolumn{2}{c|}{$^{56}$Fe/D} & \multicolumn{2}{c|}{$^{208}$Pb/D} \\
            &  ~[GeV$^2$]~  & EXT/INC & ~EMC ratio~ & EXT/INC & ~EMC ratio~ & EXT/INC & ~EMC ratio~ & EXT/INC & ~EMC ratio~   \\
    \hline
~0.220~ & 1.62 & 0.9750 & $1.0276(490)$ & 0.9730 & $0.9785(518)$ & 0.9870 & $1.0129(506)$ & 0.9950 & $1.0386(510)$ \\  
  0.247 & 1.72 & 0.9797 & $1.0111(094)$ & 0.9798 & $0.9869(093)$ & 0.9910 & $1.0112(094)$ & 0.9984 & $1.0281(095)$ \\  
  0.260 & 1.77 & 0.9820 & $1.0036(094)$ & 0.9830 & $0.9835(093)$ & 0.9930 & $1.0087(094)$ & 1.0000 & $1.0201(095)$ \\  \hline
  0.273 & 1.81 & 0.9843 & $1.0020(094)$ & 0.9856 & $0.9893(093)$ & 0.9953 & $1.0095(094)$ & 1.0019 & $1.0176(095)$ \\  
  0.287 & 1.86 & 0.9867 & $0.9956(094)$ & 0.9884 & $0.9905(094)$ & 0.9977 & $1.0045(094)$ & 1.0041 & $1.0141(095)$ \\  
  0.300 & 1.90 & 0.9890 & $0.9939(094)$ & 0.9910 & $0.9903(094)$ & 1.0000 & $1.0022(094)$ & 1.0060 & $1.0093(095)$ \\  \hline
  0.313 & 1.94 & 0.9916 & $0.9995(094)$ & 0.9939 & $0.9896(094)$ & 1.0019 & $1.0058(095)$ & 1.0086 & $1.0103(095)$ \\  
  0.327 & 1.98 & 0.9944 & $1.0033(094)$ & 0.9971 & $0.9980(094)$ & 1.0041 & $1.0074(095)$ & 1.0114 & $1.0148(095)$ \\  
  0.340 & 2.02 & 0.9970 & $1.0020(094)$ & 1.0000 & $0.9971(094)$ & 1.0060 & $1.0027(095)$ & 1.0140 & $1.0126(095)$ \\  \hline
  0.353 & 2.05 & 0.9983 & $0.9923(094)$ & 1.0010 & $0.9812(094)$ & 1.0073 & $0.9892(094)$ & 1.0153 & $1.0017(095)$ \\  
  0.367 & 2.08 & 0.9997 & $0.9887(094)$ & 1.0020 & $0.9795(094)$ & 1.0087 & $0.9859(094)$ & 1.0167 & $0.9962(095)$ \\  
  0.380 & 2.11 & 1.0010 & $0.9860(094)$ & 1.0030 & $0.9776(094)$ & 1.0100 & $0.9835(094)$ & 1.0180 & $0.9924(086)$ \\  \hline
  0.393 & 2.16 & 1.0016 & $0.9776(094)$ & 1.0040 & $0.9707(094)$ & 1.0110 & $0.9737(094)$ & 1.0187 & $0.9817(086)$ \\  
  0.407 & 2.24 & 1.0023 & $0.9933(094)$ & 1.0050 & $0.9871(094)$ & 1.0120 & $0.9854(094)$ & 1.0193 & $0.9967(095)$ \\  
  0.420 & 2.33 & 1.0030 & $0.9829(094)$ & 1.0060 & $0.9782(094)$ & 1.0130 & $0.9775(094)$ & 1.0200 & $0.9805(086)$ \\  \hline
  0.433 & 2.41 & 1.0036 & $0.9625(094)$ & 1.0063 & $0.9564(093)$ & 1.0133 & $0.9567(085)$ & 1.0200 & $0.9626(086)$ \\  
  0.447 & 2.50 & 1.0043 & $0.9612(094)$ & 1.0067 & $0.9581(094)$ & 1.0137 & $0.9534(085)$ & 1.0200 & $0.9640(086)$ \\  
  0.460 & 2.60 & 1.0050 & $0.9547(093)$ & 1.0070 & $0.9474(093)$ & 1.0140 & $0.9438(085)$ & 1.0200 & $0.9481(085)$ \\  \hline
  0.473 & 2.70 & 1.0056 & $0.9614(094)$ & 1.0076 & $0.9564(094)$ & 1.0150 & $0.9492(085)$ & 1.0203 & $0.9562(085)$ \\  
  0.487 & 2.80 & 1.0063 & $0.9460(093)$ & 1.0083 & $0.9369(093)$ & 1.0160 & $0.9331(085)$ & 1.0207 & $0.9428(085)$ \\  
  0.500 & 2.91 & 1.0070 & $0.9456(093)$ & 1.0090 & $0.9428(093)$ & 1.0170 & $0.9251(084)$ & 1.0210 & $0.9325(085)$ \\  \hline
  0.513 & 3.02 & 1.0086 & $0.9562(102)$ & 1.0106 & $0.9435(093)$ & 1.0193 & $0.9369(085)$ & 1.0231 & $0.9356(094)$ \\  
  0.527 & 3.14 & 1.0104 & $0.9457(102)$ & 1.0124 & $0.9218(093)$ & 1.0217 & $0.9176(084)$ & 1.0254 & $0.9194(093)$ \\  
  0.540 & 3.25 & 1.0120 & $0.9422(102)$ & 1.0140 & $0.9276(093)$ & 1.0240 & $0.9138(084)$ & 1.0275 & $0.9182(094)$ \\  \hline
  0.553 & 3.37 & 1.0139 & $0.9186(198)$ & 1.0159 & $0.8967(198)$ & 1.0263 & $0.8974(180)$ & 1.0296 & $0.8915(180)$ \\  
  0.580 & 3.60 & 1.0180 & $0.9427(471)$ & 1.0200 & $0.9481(482)$ & 1.0310 & $0.9360(466)$ & 1.0340 & $0.8978(426)$ \\

    \hline
    \end{tabular}
    \caption{The ratio of the EXTERNALS RC factor to that from INCLUSIVE, along with the CLAS A/D per-nucleon cross-section ratio with the EXTERNALS RCs applied.}
    \label{tab:EMCratios}
\end{table*}

\end{document}

%% file: authors.tex
\newcommand*{\JLAB}{Thomas Jefferson National Accelerator Facility, Newport News, VA 23606, USA}
\newcommand*{\LBL}{Lawrence Berkeley National Laboratory, Berkeley, CA 94720, USA}
\newcommand*{\UCR}{Department of Physics and Astronomy, University of CA, Riverside, California 92521, USA }

\author{S. Moran}
\affiliation{\UCR}
\author{M. Arratia}
\affiliation{\UCR}
\author{J. Arrington}
\affiliation{\LBL}
\author{D. Gaskell}
\affiliation{\JLAB}
\author{B. Schmookler}
\affiliation{\UCR}

%% file: EMC-RC.bbl
\begin{thebibliography}{29}%
\makeatletter
\providecommand \@ifxundefined [1]{%
 \@ifx{#1\undefined}
}%
\providecommand \@ifnum [1]{%
 \ifnum #1\expandafter \@firstoftwo
 \else \expandafter \@secondoftwo
 \fi
}%
\providecommand \@ifx [1]{%
 \ifx #1\expandafter \@firstoftwo
 \else \expandafter \@secondoftwo
 \fi
}%
\providecommand \natexlab [1]{#1}%
\providecommand \enquote  [1]{``#1''}%
\providecommand \bibnamefont  [1]{#1}%
\providecommand \bibfnamefont [1]{#1}%
\providecommand \citenamefont [1]{#1}%
\providecommand \href@noop [0]{\@secondoftwo}%
\providecommand \href [0]{\begingroup \@sanitize@url \@href}%
\providecommand \@href[1]{\@@startlink{#1}\@@href}%
\providecommand \@@href[1]{\endgroup#1\@@endlink}%
\providecommand \@sanitize@url [0]{\catcode `\\12\catcode `\$12\catcode
  `\&12\catcode `\#12\catcode `\^12\catcode `\_12\catcode `\%12\relax}%
\providecommand \@@startlink[1]{}%
\providecommand \@@endlink[0]{}%
\providecommand \url  [0]{\begingroup\@sanitize@url \@url }%
\providecommand \@url [1]{\endgroup\@href {#1}{\urlprefix }}%
\providecommand \urlprefix  [0]{URL }%
\providecommand \Eprint [0]{\href }%
\providecommand \doibase [0]{http://dx.doi.org/}%
\providecommand \selectlanguage [0]{\@gobble}%
\providecommand \bibinfo  [0]{\@secondoftwo}%
\providecommand \bibfield  [0]{\@secondoftwo}%
\providecommand \translation [1]{[#1]}%
\providecommand \BibitemOpen [0]{}%
\providecommand \bibitemStop [0]{}%
\providecommand \bibitemNoStop [0]{.\EOS\space}%
\providecommand \EOS [0]{\spacefactor3000\relax}%
\providecommand \BibitemShut  [1]{\csname bibitem#1\endcsname}%
\let\auto@bib@innerbib\@empty
\bibitem [{\citenamefont {Aubert}\ \emph {et~al.}(1983)\citenamefont {Aubert}
  \emph {et~al.}}]{EuropeanMuon:1983wih}%
  \BibitemOpen
  \bibfield  {author} {\bibinfo {author} {\bibfnamefont {J.~J.}\ \bibnamefont
  {Aubert}} \emph {et~al.} (\bibinfo {collaboration} {European Muon}),\ }\href
  {\doibase 10.1016/0370-2693(83)90437-9} {\bibfield  {journal} {\bibinfo
  {journal} {Phys. Lett. B}\ }\textbf {\bibinfo {volume} {123}},\ \bibinfo
  {pages} {275} (\bibinfo {year} {1983})}\BibitemShut {NoStop}%
\bibitem [{\citenamefont {Gomez}\ \emph {et~al.}(1994)\citenamefont {Gomez}
  \emph {et~al.}}]{Gomez:1993ri}%
  \BibitemOpen
  \bibfield  {author} {\bibinfo {author} {\bibfnamefont {J.}~\bibnamefont
  {Gomez}} \emph {et~al.},\ }\href {\doibase 10.1103/PhysRevD.49.4348}
  {\bibfield  {journal} {\bibinfo  {journal} {Phys. Rev. D}\ }\textbf {\bibinfo
  {volume} {49}},\ \bibinfo {pages} {4348} (\bibinfo {year}
  {1994})}\BibitemShut {NoStop}%
\bibitem [{\citenamefont {Dasu}\ \emph {et~al.}(1994)\citenamefont {Dasu},
  \citenamefont {deBarbaro}, \citenamefont {Bodek}, \citenamefont {Harada},
  \citenamefont {Krasny} \emph {et~al.}}]{Dasu:1993vk}%
  \BibitemOpen
  \bibfield  {author} {\bibinfo {author} {\bibfnamefont {S.}~\bibnamefont
  {Dasu}}, \bibinfo {author} {\bibfnamefont {P.}~\bibnamefont {deBarbaro}},
  \bibinfo {author} {\bibfnamefont {A.}~\bibnamefont {Bodek}}, \bibinfo
  {author} {\bibfnamefont {H.}~\bibnamefont {Harada}}, \bibinfo {author}
  {\bibfnamefont {M.}~\bibnamefont {Krasny}},  \emph {et~al.},\ }\href
  {\doibase 10.1103/PhysRevD.49.5641} {\bibfield  {journal} {\bibinfo
  {journal} {Phys. Rev. D}\ }\textbf {\bibinfo {volume} {49}},\ \bibinfo
  {pages} {5641} (\bibinfo {year} {1994})}\BibitemShut {NoStop}%
\bibitem [{\citenamefont {Seely}\ \emph {et~al.}(2009)\citenamefont {Seely}
  \emph {et~al.}}]{Seely:2009gt}%
  \BibitemOpen
  \bibfield  {author} {\bibinfo {author} {\bibfnamefont {J.}~\bibnamefont
  {Seely}} \emph {et~al.},\ }\href {\doibase 10.1103/PhysRevLett.103.202301}
  {\bibfield  {journal} {\bibinfo  {journal} {Phys. Rev. Lett.}\ }\textbf
  {\bibinfo {volume} {103}},\ \bibinfo {pages} {202301} (\bibinfo {year}
  {2009})}\BibitemShut {NoStop}%
\bibitem [{\citenamefont {Schmookler}\ \emph {et~al.}(2019)\citenamefont
  {Schmookler} \emph {et~al.}}]{CLAS:2019vsb}%
  \BibitemOpen
  \bibfield  {author} {\bibinfo {author} {\bibfnamefont {B.}~\bibnamefont
  {Schmookler}} \emph {et~al.},\ }\href {\doibase 10.1038/s41586-019-0925-9}
  {\bibfield  {journal} {\bibinfo  {journal} {Nature}\ }\textbf {\bibinfo
  {volume} {566}},\ \bibinfo {pages} {354} (\bibinfo {year}
  {2019})}\BibitemShut {NoStop}%
\bibitem [{\citenamefont {Arrington}\ \emph {et~al.}(2021)\citenamefont
  {Arrington} \emph {et~al.}}]{Arrington:2021vuu}%
  \BibitemOpen
  \bibfield  {author} {\bibinfo {author} {\bibfnamefont {J.}~\bibnamefont
  {Arrington}} \emph {et~al.},\ }\href {\doibase 10.1103/PhysRevC.104.065203}
  {\bibfield  {journal} {\bibinfo  {journal} {Phys. Rev. C}\ }\textbf {\bibinfo
  {volume} {104}},\ \bibinfo {pages} {065203} (\bibinfo {year}
  {2021})}\BibitemShut {NoStop}%
\bibitem [{\citenamefont {Abrams}\ \emph {et~al.}(2022)\citenamefont {Abrams}
  \emph {et~al.}}]{JeffersonLabHallATritium:2021usd}%
  \BibitemOpen
  \bibfield  {author} {\bibinfo {author} {\bibfnamefont {D.}~\bibnamefont
  {Abrams}} \emph {et~al.} (\bibinfo {collaboration} {Jefferson Lab Hall A
  Tritium}),\ }\href {\doibase 10.1103/PhysRevLett.128.132003} {\bibfield
  {journal} {\bibinfo  {journal} {Phys. Rev. Lett.}\ }\textbf {\bibinfo
  {volume} {128}},\ \bibinfo {pages} {132003} (\bibinfo {year} {2022})},\
  \Eprint {http://arxiv.org/abs/2104.05850} {arXiv:2104.05850 [hep-ex]}
  \BibitemShut {NoStop}%
\bibitem [{\citenamefont {Karki}\ \emph {et~al.}(2023)\citenamefont {Karki}
  \emph {et~al.}}]{HallC:2023xkj}%
  \BibitemOpen
  \bibfield  {author} {\bibinfo {author} {\bibfnamefont {A.}~\bibnamefont
  {Karki}} \emph {et~al.},\ }\href {\doibase 10.1103/PhysRevC.108.035201}
  {\bibfield  {journal} {\bibinfo  {journal} {Phys. Rev. C}\ }\textbf {\bibinfo
  {volume} {108}},\ \bibinfo {pages} {035201} (\bibinfo {year}
  {2023})}\BibitemShut {NoStop}%
\bibitem [{\citenamefont {Malace}\ \emph {et~al.}(2014)\citenamefont {Malace},
  \citenamefont {Gaskell}, \citenamefont {Higinbotham},\ and\ \citenamefont
  {Cloet}}]{Malace:2014uea}%
  \BibitemOpen
  \bibfield  {author} {\bibinfo {author} {\bibfnamefont {S.}~\bibnamefont
  {Malace}}, \bibinfo {author} {\bibfnamefont {D.}~\bibnamefont {Gaskell}},
  \bibinfo {author} {\bibfnamefont {D.~W.}\ \bibnamefont {Higinbotham}}, \ and\
  \bibinfo {author} {\bibfnamefont {I.}~\bibnamefont {Cloet}},\ }\href
  {\doibase 10.1142/S0218301314300136} {\bibfield  {journal} {\bibinfo
  {journal} {Int. J. Mod. Phys. E}\ }\textbf {\bibinfo {volume} {23}},\
  \bibinfo {pages} {1430013} (\bibinfo {year} {2014})}\BibitemShut {NoStop}%
\bibitem [{\citenamefont {Arrington}(2016)}]{Arrington:2015wja}%
  \BibitemOpen
  \bibfield  {author} {\bibinfo {author} {\bibfnamefont {J.}~\bibnamefont
  {Arrington}},\ }\href {\doibase 10.1051/epjconf/201611301011} {\bibfield
  {journal} {\bibinfo  {journal} {EPJ Web Conf.}\ }\textbf {\bibinfo {volume}
  {113}},\ \bibinfo {pages} {01011} (\bibinfo {year} {2016})},\ \Eprint
  {http://arxiv.org/abs/1508.05042} {arXiv:1508.05042 [nucl-ex]} \BibitemShut
  {NoStop}%
\bibitem [{\citenamefont {Hen}\ \emph {et~al.}(2017)\citenamefont {Hen},
  \citenamefont {Miller}, \citenamefont {Piasetzky},\ and\ \citenamefont
  {Weinstein}}]{Hen:2016kwk}%
  \BibitemOpen
  \bibfield  {author} {\bibinfo {author} {\bibfnamefont {O.}~\bibnamefont
  {Hen}}, \bibinfo {author} {\bibfnamefont {G.~A.}\ \bibnamefont {Miller}},
  \bibinfo {author} {\bibfnamefont {E.}~\bibnamefont {Piasetzky}}, \ and\
  \bibinfo {author} {\bibfnamefont {L.~B.}\ \bibnamefont {Weinstein}},\ }\href
  {\doibase 10.1103/RevModPhys.89.045002} {\bibfield  {journal} {\bibinfo
  {journal} {Rev. Mod. Phys.}\ }\textbf {\bibinfo {volume} {89}},\ \bibinfo
  {pages} {045002} (\bibinfo {year} {2017})},\ \Eprint
  {http://arxiv.org/abs/1611.09748} {arXiv:1611.09748 [nucl-ex]} \BibitemShut
  {NoStop}%
\bibitem [{\citenamefont {Clo\"et}\ \emph {et~al.}(2019)\citenamefont {Clo\"et}
  \emph {et~al.}}]{Cloet:2019mql}%
  \BibitemOpen
  \bibfield  {author} {\bibinfo {author} {\bibfnamefont {I.~C.}\ \bibnamefont
  {Clo\"et}} \emph {et~al.},\ }\href {\doibase 10.1088/1361-6471/ab2731}
  {\bibfield  {journal} {\bibinfo  {journal} {J. Phys. G}\ }\textbf {\bibinfo
  {volume} {46}},\ \bibinfo {pages} {093001} (\bibinfo {year}
  {2019})}\BibitemShut {NoStop}%
\bibitem [{\citenamefont {Arrington}\ \emph {et~al.}(2022)\citenamefont
  {Arrington}, \citenamefont {Fomin},\ and\ \citenamefont
  {Schmidt}}]{Arrington:2022sov}%
  \BibitemOpen
  \bibfield  {author} {\bibinfo {author} {\bibfnamefont {J.}~\bibnamefont
  {Arrington}}, \bibinfo {author} {\bibfnamefont {N.}~\bibnamefont {Fomin}}, \
  and\ \bibinfo {author} {\bibfnamefont {A.}~\bibnamefont {Schmidt}},\
  }\href@noop {} {\bibfield  {journal} {\bibinfo  {journal} {Ann. Rev. Nucl.
  Part. Sci.}\ }\textbf {\bibinfo {volume} {72}},\ \bibinfo {pages} {307}
  (\bibinfo {year} {2022})}\BibitemShut {NoStop}%
\bibitem [{\citenamefont {Fomin}\ \emph {et~al.}(2012)\citenamefont {Fomin}
  \emph {et~al.}}]{Fomin:2011ng}%
  \BibitemOpen
  \bibfield  {author} {\bibinfo {author} {\bibfnamefont {N.}~\bibnamefont
  {Fomin}} \emph {et~al.},\ }\href {\doibase 10.1103/PhysRevLett.108.092502}
  {\bibfield  {journal} {\bibinfo  {journal} {Phys. Rev. Lett.}\ }\textbf
  {\bibinfo {volume} {108}},\ \bibinfo {pages} {092502} (\bibinfo {year}
  {2012})}\BibitemShut {NoStop}%
\bibitem [{\citenamefont {Arrington}\ \emph {et~al.}(2012)\citenamefont
  {Arrington} \emph {et~al.}}]{Arrington:2012ax}%
  \BibitemOpen
  \bibfield  {author} {\bibinfo {author} {\bibfnamefont {J.}~\bibnamefont
  {Arrington}} \emph {et~al.},\ }\href {\doibase 10.1103/PhysRevC.86.065204}
  {\bibfield  {journal} {\bibinfo  {journal} {Phys. Rev. C}\ }\textbf {\bibinfo
  {volume} {86}},\ \bibinfo {pages} {065204} (\bibinfo {year}
  {2012})}\BibitemShut {NoStop}%
\bibitem [{\citenamefont {Tsai}(1971)}]{Tsai:1971qi}%
  \BibitemOpen
  \bibfield  {author} {\bibinfo {author} {\bibfnamefont {Y.~S.}\ \bibnamefont
  {Tsai}},\ }\href@noop {} {} (\bibinfo {year} {1971}),\ \bibinfo {note}
  {{SLAC-PUB-848}}\BibitemShut {NoStop}%
\bibitem [{\citenamefont {Sargsian}(1990)}]{inclusive}%
  \BibitemOpen
  \bibfield  {author} {\bibinfo {author} {\bibfnamefont {M.}~\bibnamefont
  {Sargsian}},\ }\href@noop {} {\enquote {\bibinfo {title} {{Computer Code for
  Inclusive (e,e') Electro-production Reactions and Radiative Corrections}},}\
  } (\bibinfo {year} {1990}),\ \bibinfo {note} {{CLAS Note 90-007,
  \url{https://www.jlab.org/Hall-B/notes/clas_notes-all.html}}}\BibitemShut
  {NoStop}%
\bibitem [{sup(2023)}]{supplemental}%
  \BibitemOpen
  \href@noop {} {\enquote {\bibinfo {title} {{Supplemental Material}},}\ }
  (\bibinfo {year} {2023})\BibitemShut {NoStop}%
\bibitem [{\citenamefont {Hakobyan}\ \emph {et~al.}(2008)\citenamefont
  {Hakobyan} \emph {et~al.}}]{Hakobyan:2008zz}%
  \BibitemOpen
  \bibfield  {author} {\bibinfo {author} {\bibfnamefont {H.}~\bibnamefont
  {Hakobyan}} \emph {et~al.},\ }\href {\doibase 10.1016/j.nima.2008.04.055}
  {\bibfield  {journal} {\bibinfo  {journal} {Nucl. Instrum. Meth. A}\ }\textbf
  {\bibinfo {volume} {592}},\ \bibinfo {pages} {218} (\bibinfo {year}
  {2008})}\BibitemShut {NoStop}%
\bibitem [{\citenamefont {Frankfurt}\ and\ \citenamefont
  {Strikman}(1988)}]{frankfurt88}%
  \BibitemOpen
  \bibfield  {author} {\bibinfo {author} {\bibfnamefont {L.}~\bibnamefont
  {Frankfurt}}\ and\ \bibinfo {author} {\bibfnamefont {M.}~\bibnamefont
  {Strikman}},\ }\href {\doibase 10.1016/0370-1573(88)90179-2} {\bibfield
  {journal} {\bibinfo  {journal} {Phys. Rept.}\ }\textbf {\bibinfo {volume}
  {160}},\ \bibinfo {pages} {235} (\bibinfo {year} {1988})}\BibitemShut
  {NoStop}%
\bibitem [{\citenamefont {Sargsian}\ \emph {et~al.}(2003)\citenamefont
  {Sargsian} \emph {et~al.}}]{sargsian03}%
  \BibitemOpen
  \bibfield  {author} {\bibinfo {author} {\bibfnamefont {M.~M.}\ \bibnamefont
  {Sargsian}} \emph {et~al.},\ }\href {\doibase 10.1088/0954-3899/29/3/201}
  {\bibfield  {journal} {\bibinfo  {journal} {J. Phys.}\ }\textbf {\bibinfo
  {volume} {G29}},\ \bibinfo {pages} {R1} (\bibinfo {year} {2003})}\BibitemShut
  {NoStop}%
\bibitem [{\citenamefont {Fomin}\ \emph {et~al.}(2017)\citenamefont {Fomin},
  \citenamefont {Higinbotham}, \citenamefont {Sargsian},\ and\ \citenamefont
  {Solvignon}}]{fomin17}%
  \BibitemOpen
  \bibfield  {author} {\bibinfo {author} {\bibfnamefont {N.}~\bibnamefont
  {Fomin}}, \bibinfo {author} {\bibfnamefont {D.}~\bibnamefont {Higinbotham}},
  \bibinfo {author} {\bibfnamefont {M.}~\bibnamefont {Sargsian}}, \ and\
  \bibinfo {author} {\bibfnamefont {P.}~\bibnamefont {Solvignon}},\ }\href
  {\doibase 10.1146/annurev-nucl-102115-044939} {\bibfield  {journal} {\bibinfo
   {journal} {Ann. Rev. Nucl. Part. Sci.}\ }\textbf {\bibinfo {volume} {67}},\
  \bibinfo {pages} {129} (\bibinfo {year} {2017})}\BibitemShut {NoStop}%
\bibitem [{\citenamefont {Frankfurt}\ \emph {et~al.}(1993)\citenamefont
  {Frankfurt}, \citenamefont {Strikman}, \citenamefont {Day},\ and\
  \citenamefont {Sargsyan}}]{frankfurt93}%
  \BibitemOpen
  \bibfield  {author} {\bibinfo {author} {\bibfnamefont {L.~L.}\ \bibnamefont
  {Frankfurt}}, \bibinfo {author} {\bibfnamefont {M.~I.}\ \bibnamefont
  {Strikman}}, \bibinfo {author} {\bibfnamefont {D.~B.}\ \bibnamefont {Day}}, \
  and\ \bibinfo {author} {\bibfnamefont {M.}~\bibnamefont {Sargsyan}},\ }\href
  {\doibase 10.1103/PhysRevC.48.2451} {\bibfield  {journal} {\bibinfo
  {journal} {Phys. Rev. C}\ }\textbf {\bibinfo {volume} {48}},\ \bibinfo
  {pages} {2451} (\bibinfo {year} {1993})}\BibitemShut {NoStop}%
\bibitem [{\citenamefont {Li}\ \emph {et~al.}(2022)\citenamefont {Li} \emph
  {et~al.}}]{Li:2022fhh}%
  \BibitemOpen
  \bibfield  {author} {\bibinfo {author} {\bibfnamefont {S.}~\bibnamefont {Li}}
  \emph {et~al.},\ }\href {\doibase 10.1038/s41586-022-05007-2} {\bibfield
  {journal} {\bibinfo  {journal} {Nature}\ }\textbf {\bibinfo {volume} {609}},\
  \bibinfo {pages} {41} (\bibinfo {year} {2022})},\ \Eprint
  {http://arxiv.org/abs/2210.04189} {arXiv:2210.04189 [nucl-ex]} \BibitemShut
  {NoStop}%
\bibitem [{\citenamefont {Moran}\ \emph {et~al.}(2022)\citenamefont {Moran}
  \emph {et~al.}}]{CLAS:2021jhm}%
  \BibitemOpen
  \bibfield  {author} {\bibinfo {author} {\bibfnamefont {S.}~\bibnamefont
  {Moran}} \emph {et~al.} (\bibinfo {collaboration} {CLAS}),\ }\href {\doibase
  10.1103/PhysRevC.105.015201} {\bibfield  {journal} {\bibinfo  {journal}
  {Phys. Rev. C}\ }\textbf {\bibinfo {volume} {105}},\ \bibinfo {pages}
  {015201} (\bibinfo {year} {2022})},\ \Eprint
  {http://arxiv.org/abs/2109.09951} {arXiv:2109.09951 [nucl-ex]} \BibitemShut
  {NoStop}%
\bibitem [{\citenamefont {Chetry}\ \emph {et~al.}(2023)\citenamefont {Chetry}
  \emph {et~al.}}]{CLAS:2022oux}%
  \BibitemOpen
  \bibfield  {author} {\bibinfo {author} {\bibfnamefont {T.}~\bibnamefont
  {Chetry}} \emph {et~al.} (\bibinfo {collaboration} {CLAS}),\ }\href {\doibase
  10.1103/PhysRevLett.130.142301} {\bibfield  {journal} {\bibinfo  {journal}
  {Phys. Rev. Lett.}\ }\textbf {\bibinfo {volume} {130}},\ \bibinfo {pages}
  {142301} (\bibinfo {year} {2023})}\BibitemShut {NoStop}%
\bibitem [{\citenamefont {Airapetian}\ \emph {et~al.}(2001)\citenamefont
  {Airapetian} \emph {et~al.}}]{HERMES:2000ytc}%
  \BibitemOpen
  \bibfield  {author} {\bibinfo {author} {\bibfnamefont {A.}~\bibnamefont
  {Airapetian}} \emph {et~al.} (\bibinfo {collaboration} {HERMES}),\ }\href
  {\doibase 10.1007/s100520100697} {\bibfield  {journal} {\bibinfo  {journal}
  {Eur. Phys. J. C}\ }\textbf {\bibinfo {volume} {20}},\ \bibinfo {pages} {479}
  (\bibinfo {year} {2001})},\ \Eprint {http://arxiv.org/abs/hep-ex/0012049}
  {arXiv:hep-ex/0012049} \BibitemShut {NoStop}%
\bibitem [{\citenamefont {Airapetian}\ \emph {et~al.}(2011)\citenamefont
  {Airapetian} \emph {et~al.}}]{HERMES:2011qjb}%
  \BibitemOpen
  \bibfield  {author} {\bibinfo {author} {\bibfnamefont {A.}~\bibnamefont
  {Airapetian}} \emph {et~al.} (\bibinfo {collaboration} {HERMES}),\ }\href
  {\doibase 10.1140/epja/i2011-11113-5} {\bibfield  {journal} {\bibinfo
  {journal} {Eur. Phys. J. A}\ }\textbf {\bibinfo {volume} {47}},\ \bibinfo
  {pages} {113} (\bibinfo {year} {2011})},\ \Eprint
  {http://arxiv.org/abs/1107.3496} {arXiv:1107.3496 [hep-ex]} \BibitemShut
  {NoStop}%
\bibitem [{\citenamefont {Bosted}\ and\ \citenamefont
  {Mamyan}(2012)}]{Bosted:2012qc}%
  \BibitemOpen
  \bibfield  {author} {\bibinfo {author} {\bibfnamefont {P.}~\bibnamefont
  {Bosted}}\ and\ \bibinfo {author} {\bibfnamefont {V.}~\bibnamefont
  {Mamyan}},\ }\href@noop {} {\bibfield  {journal} {\bibinfo  {journal}
  {arXiv:1203.2262}\ } (\bibinfo {year} {2012})}\BibitemShut {NoStop}%
\end{thebibliography}%
